%% file: ANA-EXOT-2019-27-PAPER.tex
\pdfoutput=1
\documentclass[PAPER, atlasdraft=false, UKenglish, coverpage=false, cernpreprint=true, texmf, orcidlogo]{atlasdoc}
\usepackage[subcaption]{atlaspackage}
\usepackage{atlasbiblatex}

\usepackage{atlasphysics}
\usepackage{amsmath,booktabs}
\usepackage{siunitx}
\usepackage{xspace}
\usepackage{multirow}
\usepackage{slashed}
\usepackage{rotating}
\graphicspath{{logos/}{figures/}}

\usepackage{ANA-EXOT-2019-27-PAPER-defs}

\LEcontact{Bob Kowalewski bob.kowalewski@cern.ch}

\AtlasTitle{Search for new particles in events with a hadronically decaying W or Z boson and large missing transverse momentum at $\sqrt{s}$ = 13~TeV using the ATLAS detector}

\AtlasAbstract{%
A search is presented for new particles produced in proton--proton collisions at a centre-of-mass energy of \SI{13}{TeV} that result in final states comprising a massive vector ($W$ or $Z$) boson that decays hadronically and large missing transverse momentum. The data sample was collected with the ATLAS experiment at the Large Hadron Collider from 2015 to 2018 and corresponds to an integrated luminosity of 140~fb$^{-1}$. No significant excess over the Standard Model expectation is observed. Model-independent 95\% confidence-level limits on the visible cross-section that range from 0.3~fb to 79.5~fb are obtained for non-Standard-Model processes. Exclusion limits are also presented for models with axion-like particles, for two-Higgs-doublet models with a pseudo-scalar mediator between the Standard Model and the dark sector, for the invisible decay of the Higgs boson and for pair-produced weakly interacting dark matter candidates.
}

\author{The ATLAS Collaboration}

\PreprintIdNumber{CERN-EP-2024-128}

\AtlasJournalRef{\JHEP 11 (2024) 126}
\AtlasDOI{10.1007/JHEP11(2024)126}

\hypersetup{pdftitle={ATLAS document},pdfauthor={The ATLAS Collaboration}}

\begin{document}

\maketitle

\tableofcontents

%
%

%

\section{Introduction}

Many astrophysics measurements~\cite{Bertone:2004pz,Feng:2010gw,Porter:2011nv} have indicated the presence of a non-baryonic dark matter (DM) component in the Universe, but the nature of such DM, and how it interacts with normal matter, remain unknown. The discovery of DM particles and their interactions with the Standard Model (SM) particles is one of the main quests in particle physics. At the Large Hadron Collider (LHC)~\cite{Evans:2008zzb}, DM particles may be produced in proton--proton collisions and would traverse the detector without interacting. The observational signature of these non-interacting DM particles would be a large missing transverse momentum (\met) recoiling against visible SM particles. Searches for DM candidates can be performed in different final states distinguished by the detected SM particle(s) $X$ ($\MET+X$). In this paper, a search for new particles is performed in final states with a hadronically decaying $V$ ($W$ or $Z$) boson, missing transverse momentum and no leptons ($\MET+V$).

The $\MET+V$ signature can be used to explore new particles proposed in many theories beyond the SM such as axion-like particles, two-Higgs-doublet models with a pseudoscalar mediator between the Standard Model and the dark sector, the invisible decay of the Higgs boson and pair-produced weakly interacting dark matter candidates.

\begin{figure}[tbp]
\centering
\begin{subfigure}[t]{0.4\textwidth}
\centering
\includegraphics[width = 0.99\textwidth]{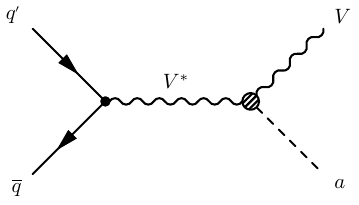}
\caption{}
\end{subfigure}
\begin{subfigure}[t]{0.4\textwidth}
\centering
\includegraphics[width = 0.99\textwidth]{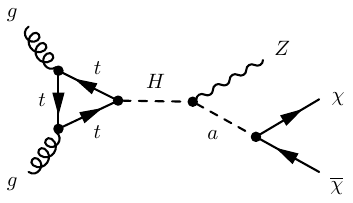}
\caption{}
\end{subfigure}
\begin{subfigure}[t]{0.4\textwidth}
\centering
\includegraphics[width = 0.99\textwidth]{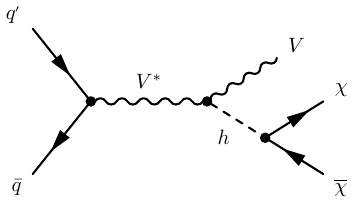}
\caption{}
\end{subfigure}
\begin{subfigure}[t]{0.4\textwidth}
\centering
\includegraphics[width = 0.99\textwidth]{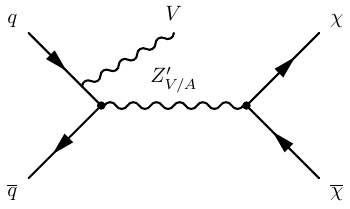}
\caption{}
\end{subfigure}
\caption{Feynman diagrams for: (a) the production of an ALP $a$ in association with a $V$ boson, (b) the $\MET+Z$ signature production in the 2HDM model with the pseudo-scalar mediator $a$, (c) the Higgs boson production in association with a $V$ boson in which the Higgs boson decays into DM candidates $\chi$, and (d) the pair production of weakly interacting massive particles $\chi$ through a mediator $Z’_{V/A}$ with the emission of a $V$ as initial state radiation.}
\label{fig:signal}
\end{figure}

Axion-like particles (ALPs) are weakly interacting pseudo-scalars. Axions were proposed to solve the charge-parity (CP) symmetry problem in QCD interactions through adding a global U(1) symmetry to the SM~\cite{PhysRevLett.38.1440}. The ALPs model is described with an effective ﬁeld theory (EFT) that extends the SM Lagrangian using effective operators for the interactions between ordinary matter and the ALPs ($a$)~\cite{Brivio:2017ije}. The new scale associated with the physics of the ALP (an effective scale ($f_{a}$)) regulates the dimension-5 operators built from the SM fields and the ALPs. The analysis reported here uses, for the first time, a model with the ALP generated in association with a $V$ boson as illustrated in Figure~\ref{fig:signal} (a).

Higgs-related DM models can be tested extensively at the LHC energy frontier and are well motivated after the discovery of the Higgs boson by the ATLAS and CMS Collaborations in 2012~\cite{HIGG-2012-27, CMS-HIG-12-028}. A two-Higgs-doublet model with a pseudo-scalar mediator (2HDM+$a$) between the SM and the dark sector~\cite{Bauer:2017ota} is hence also explored. In this model, the SM Higgs sector is extended by five physical scalar states in addition to the SM-like Higgs boson $h$: a scalar ($H$), pseudo-scalar ($A$), two charged Higgs bosons ($H^{\pm}$), and the pseudo-scalar mediator ($a$) that couples to DM particles. The phenomenology of the model is fully determined by 14 independent parameters:  the masses of the Higgs bosons ($m_H$, $m_A$, and $m_{H^{\pm}}$), the mass of the mediator ($m_{a}$), the mass of the DM particle ($m_{\chi}$), the Yukawa coupling strength between the mediator and the DM particle ($g_{\chi}$), the electroweak vacuum expectation value (VEV), the ratio of the VEVs of the two Higgs doublets ($\tan\beta$), the mixing angles of the CP-even ($\alpha$) and CP-odd ($\theta$) weak eigenstates, the quartic coupling ($\lambda_{3}$) of the pure 2HDM potential term and the two quartic couplings of the potential terms connecting the doublet ($\lambda_{\textrm{P1}}$) and singlet fields ($\lambda_{\textrm{P2}}$). Only the production in association with the $Z$ boson, as shown in Figure~\ref{fig:signal} (b), is considered for this search, because it can be resonantly produced through the $H$ boson and due to the small interaction cross-section when considering the associated production with the $W$ boson. Results from other ATLAS analyses with complementary final states are combined in Ref.~\cite{EXOT-2018-64}.

If kinematically allowed, DM particles may be produced in Higgs boson decays and lead to an invisible final state. In the SM, the Higgs boson decay into undetectable final states occurs via the $ZZ^{*} \rightarrow 4 \nu$ process and has a branching fraction below 0.1\%. This process can be probed in $\MET+V$ final states through $Vh$ associated production, as shown in Figure~\ref{fig:signal} (c). In addition, the production of a Higgs boson through gluon--gluon fusion, vector boson fusion and in association with a top-quark pair ($t\bar{t}$) considered in this search to be consistent with the approach of other ATLAS analyses that probe similar models~\cite{HIGG-2021-05}. The most recent experimental results from the ATLAS and CMS Collaborations have set observed (expected) upper limits at \SI{95}{\percent} confidence level on the branching ratio of the Higgs boson decaying invisibly of 0.107 (0.077)~\cite{HIGG-2021-05} and 0.15 (0.08)~\cite{CMS-HIG-21-007}, respectively.

Weakly interacting massive particles (WIMPs)~\cite{Bertone:2017adx} can naturally produce the right thermal relic density of the universe through the freeze-out mechanism. In this paper, as a minimal extension of the SM, a simplified DM model~\cite{LHCNewPhysicsWorkingGroup:2011mji} is used to describe the production of WIMPs through the exchange of a vector boson mediator ($Z'_{V}$) or axial-vector boson mediator ($Z'_{A}$)  in the $s$-channel. The model postulates a Dirac fermion particle $\chi$ and has five free parameters~\cite{Abercrombie:2015wmb}: the mass of the Dirac fermion particle ($m_\chi$), the mediator mass ($m_{Z'})$ and the mediator couplings to quarks ($g_{q}$), to all leptons ($g_{l}$), and to DM particles ($g_{\chi}$). In Figure~\ref{fig:signal} (d) one possible Feynman diagram representing this interaction is shown, where a $V$ boson is radiated in the initial state. The $\MET+X$ signatures have been explored by the ATLAS and CMS Collaborations for the cases where $X$ is a jet~\cite{EXOT-2018-06,CMS-EXO-20-004}, photon~\cite{EXOT-2018-63,CMS-EXO-16-053}, hadronically decaying $V$~\cite{EXOT-2016-23,CMS-EXO-20-004} or leptonically decaying $Z$ boson~\cite{HIGG-2018-26,CMS-EXO-19-003}, using the proton--proton collision data collected during the LHC Run~2 (2015 -- 2018). The ATLAS $\MET+\text{jet}$ search sets the strongest bounds in this model, excluding mediator masses up to \SI{2.1}{\TeV} for DM masses of \SI{1}{GeV} assuming SM initial state radiation production rates.

This paper presents a search for an excess over the SM prediction in the $\MET+V$ signature using proton--proton collision data collected at a centre-of-mass energy of \SI{13}{\TeV} by the ATLAS experiment at the LHC. The data sample corresponds to a total integrated luminosity of 140 \ifb~\cite{DAPR-2021-01}, recorded in the period from 2015 to 2018. The sensitivity of this search is enhanced compared with the previous search, which used a partial Run~2 data sample of 36.1~\ifb~\cite{EXOT-2016-23}, due to several factors: the increased size of the data and simulated event samples; the use of improved selection criteria for reconstructed objects and reduced associated uncertainties, and improvements in the event selection.

Different analysis strategies are used depending on the boost of the $V$ boson in the final state. At low transverse momentum (\pt), the hadrons from the $V$ boson decay can be reconstructed into two well-separated jets with a radius parameter \(R=0.4\) (referred to as the resolved topology). The invariant mass of the dijet system allows discrimination between the dijet background and the resonant decay of a $V$ boson. At higher $V$ transverse momentum, it becomes less likely that the $V$ decay products are identified as individual jets in the resolved topology. In this case, the strategy is based on the identification of large-radius jets in the event, with \(R=1.0\), and the use of jet substructure quantities to select jets with an internal structure consistent with the decay of a $W$ or $Z$ boson (referred to as the merged topology).

The dominant SM background processes considered in this search include $Z(\rightarrow \nu\nu)$+jets, $W(\rightarrow \tau \nu)$+jets with the $\tau$-lepton decaying hadronically or into an unidentified charged lepton and a neutrino, $W(\rightarrow e(\mu)\nu)$+jets where the charged lepton is not identified, $Z(\rightarrow \tau\tau)$+jets with each $\tau$-lepton decaying hadronically or into unidentified charged leptons, $t\bar{t}$ and diboson production. Smaller background processes arise from single top-quark production, triboson production, Higgs boson production in association with a $V$ boson ($Vh$), $t\bar{t}$ production in association with a vector boson ($t\bar{t}V$), and multijet production. The analysis strategy defines three signal regions and twelve control regions enriched in the dominant background contributions such as $V+$jets and $t\bar{t}$.
A simultaneous likelihood fit in all signal and control regions is used to extract the background and signal estimates using the missing transverse momentum as the discriminating variable. The results are interpreted in the context of ALPs, 2HDM+$a$, the invisible decay of the Higgs boson and simplified dark matter models.

The paper is organised as follows. A brief introduction to the ATLAS detector is given in Section~\ref{sec:detector}. The data and simulated Monte Carlo samples are described in Section~\ref{sec:data}. The algorithms for the reconstruction and identification of final state particles are summarized in Section~\ref{sec:reconstruction}. Section~\ref{sec:selection} describes the criteria for the selection of signal event candidates. The experimental and theoretical systematic uncertainties are discussed, along with the statistical analysis and results, in Section~\ref{sec:result}, with their interpretation presented in Section~\ref{sec:limit}. Concluding remarks are given in Section~\ref{sec:summary}.


%

\section{ATLAS detector}
\label{sec:detector}

The ATLAS detector~\cite{PERF-2007-01} is a general-purpose detector designed to precisely measure the properties of all particles emerging from proton--proton collisions and covers nearly the entire solid angle around the collision point\footnote{ATLAS uses a right-handed coordinate system with its origin at the nominal interaction point (IP) in the centre of the detector and the \(z\)-axis along the beam pipe. The \(x\)-axis points from the IP to the centre of the LHC ring, and the \(y\)-axis points upwards. Polar coordinates \((r,\phi)\) are used in the transverse plane, \(\phi\) being the azimuthal angle around the \(z\)-axis. The pseudorapidity is defined in terms of the polar angle \(\theta\) as \(\eta = -\ln \tan(\theta/2)\) and is equal to the rapidity $ y = \frac{1}{2} \ln \left( \frac{E + p_z c}{E - p_z c} \right) $ in the relativistic limit. Angular distance is measured in units of \(\Delta R_{y} \equiv \sqrt{(\Delta y)^{2} + (\Delta\phi)^{2}}\).}. It consists of the Inner Detector (ID) surrounded by a thin superconducting solenoid that provides a \qty{2}{\tesla} axial magnetic field, the calorimeters and the Muon Spectrometer (MS).

The ID system is designed to measure the direction, momentum and charge of electrically charged particles produced in the collisions and to precisely determine the locations of primary and secondary vertices. It has full coverage in a pseudorapidity range $|\eta| < 2.5$ and consists of a silicon pixel detector, a silicon microstrip detector and a transition radiation tracker, arranged in a coaxial geometry around the beam axis.
The
innermost silicon pixel layer was added to the ID before the start of data-taking in 2015~\cite{ATLAS-TDR-19,PIX-2018-001}.

The calorimeters lie outside of the ID and measure energies of particles
in the pseudorapidity range
$|\eta| < 4.9$. The main components of the calorimeter system are the electromagnetic calorimeter, which measures the energy of electrons and photons, and the hadronic calorimeter, which measures
energy deposited by hadrons.

The MS is the outermost detector, designed to measure the deflection of muons in a magnetic field generated by three large superconducting toroid magnets with eight coils each. The field integral of the toroids ranges between \num{2.0} and \qty{6.0}{\tesla\metre} across most of the detector. The MS also provides a stand-alone muon trigger. Muons can be measured in the pseudorapidity range $|\eta| < 2.7$, and triggered in the pseudorapidity range $|\eta| < 2.4$.

Events are selected in real-time using a two-level trigger system~\cite{TRIG-2016-01}. The first level provides a hardware-based trigger decision, while the second level makes a trigger decision based on software algorithms.

An extensive software suite~\cite{SOFT-2022-02} is used in data simulation, in the reconstruction and analysis of real and simulated data, in detector operations, and in the trigger and data acquisition systems of the experiment.


%

\section{Data and Monte Carlo simulated events}
\label{sec:data}

The analysis uses proton--proton collision data at $\sqrt{s} = \SI{13}{\TeV}$ recorded by the ATLAS detector during the data-taking period from 2015 to 2018. Only data in which all subdetectors were functioning well are used. Standard detector quality criteria~\cite{DAPR-2018-01} are applied to reduce the impact of instrumental noise and out-of-time calorimeter energy deposits from cosmic rays and beam backgrounds. The data sample was collected with a bunch crossing interval (bunch spacing) of \SI{25}{\ns} and corresponds to an integrated luminosity of \SI{140}{\ifb}, with an uncertainty of \SI{0.83}{\percent}~\cite{DAPR-2021-01}, obtained using the LUCID-2~\cite{LUCID2} detector for the primary luminosity measurements, complemented by measurements using the inner detector and calorimeters.

Monte Carlo (MC) generators are used to simulate the signal and background events needed to compute the detector efficiency and predict the SM background contributions. These samples of events, together with samples generated with alternative generators, are also used to estimate the systematic uncertainties.  All simulated event samples were produced with a detailed detector simulation based on the $\GEANT$~\cite{Agostinelli:2002hh} package and employed the same particle reconstruction algorithms used on data.   Additional proton--proton interactions in the same or neighbouring bunch-crossings (pile-up) were included in the simulation process. These pile-up events were generated using the {\PYTHIA}~8.186~\cite{Sjostrand:2007gs} package with the A3 set of tuned parameters (tune)~\cite{ATL-PHYS-PUB-2016-017} and the NNPDF2.3LO parton distribution function (PDF) set~\cite{Ball:2012cx}, and are corrected to match the distribution of the average number of interactions per bunch crossing observed in the data. For all simulated event samples, except the samples generated with \SHERPA~\cite{sherpa,Cascioli:2011va,Gleisberg:2008fv}, the EvtGen 1.2.0~\cite{EvtGen} package was used to simulate the decays of $b$- and $c$-hadrons.  Compared with the previous publication~\cite{EXOT-2016-23}, the numbers of simulated events was increased substantially to reduce the statistical uncertainties arising from these samples. A summary of MC samples is shown in Table~\ref{tab:mcgen}.

\addtolength{\tabcolsep}{-2pt}
\begin{table}[tbp]
\centering{\fontsize{7}{7.9}\selectfont
\caption{The list of the Monte Carlo generators used to simulate the signal and background events.\label{tab:mcgen}}
\begin{tabular}{llllll}
\toprule
\midrule
Process & Matrix element generator & Matrix element PDF & Parton shower & Underlying event & Normalisation \\
& & & and hadronisation & model tune & order\\
\midrule
\multicolumn{6}{l}{Signal processes} \\
\midrule
Axion-like particles & \MGNLO~\cite{Alwall:2014hca}& NNPDF3.0NLO~\cite{Ball:2014uwa} & \PYTHIA 8.235~\cite{Sjostrand:2014zea} &A14~\cite{ATL-PHYS-PUB-2014-021} &NLO~\cite{Backovic:2015soa}  \\
2HDM with a pseudo-scalar & \MGNLO& NNPDF3.0NLO & \PYTHIA 8.306 &A14 &LO  \\
Invisibly decaying Higgs boson & \POWHEG-Box v2~\cite{Frixione:2007nw}& PDF4LHC15NLO~\cite{Butterworth:2015oua} & \PYTHIA 8.212 &AZNLO~\cite{STDM-2012-23} &NLO  \\
Simplified Dark Matter model & \MGNLO& NNPDF3.0NLO & \PYTHIA 8.235 &A14 &NLO~\cite{Backovic:2015soa}  \\
\midrule
\multicolumn{6}{l}{Dominant background processes} \\
\midrule
$Z$+jets & \SHERPA 2.2.11~\cite{sherpa,Cascioli:2011va,Gleisberg:2008fv} & NNPDF3.0NNLO &  \SHERPA 2.2.11 & \SHERPA Default & NNLO    \\
$W$+jets & \SHERPA 2.2.11& NNPDF3.0NNLO & \SHERPA 2.2.11 & \SHERPA Default & NNLO  \\
\multirow{2}{*}{$t\bar{t}$} &\POWHEG-Box v2~\cite{Frixione:2007nw,Nason:2004rx,Frixione:2007vw}& \multirow{2}{*}{NNPDF3.0NLO} & \multirow{2}{*}{\PYTHIA 8.230} & \multirow{2}{*}{A14} & \multirow{2}{*}{NNLO+NNLL~\cite{Czakon:2011xx}} \\
& \MGNLO& & & & \\
\multirow{2}{*}{Diboson} & \SHERPA 2.2.1 & \multirow{2}{*}{NNPDF3.0NNLO} & \SHERPA 2.2.1  & \multirow{2}{*}{\SHERPA Default} & \multirow{2}{*}{NLO}  \\
&  \SHERPA 2.2.2&  &  \SHERPA 2.2.2 & &   \\
\midrule
\multicolumn{6}{l}{Smaller background processes} \\
\midrule
Single top-quarks& \POWHEG-Box v2~\cite{Re:2010bp, Alioli:2009je}& NNPDF3.0NLO &\PYTHIA 8.230 &A14 & NLO \\
Triboson & \SHERPA 2.2.2& NNPDF3.0NNLO &  \SHERPA 2.2.2 & \SHERPA Default & NLO  \\
\multirow{2}{*}{$Vh$} & \multirow{2}{*}{\POWHEG-Box v2} & \multirow{2}{*}{NNPDF3.0NLO} & \multirow{2}{*}{\PYTHIA 8.186} & \multirow{2}{*}{AZNLO} & NNLO(QCD)/  \\
& & & & & NLO (EW)\\
$t\bar{t}V$ & \MGNLO& NNPDF3.0NLO & \PYTHIA 8.210 &A14 &NLO  \\
\midrule
\bottomrule
\end{tabular}}
\end{table}
\addtolength{\tabcolsep}{2pt}

Samples of ALP production in association with a $V$ boson are generated by suppressing processes that couple the ALP to photons, gluons or the Higgs boson. The ALP mass is set to \SI{1}{\MeV} and ALP couplings to the $V$ boson ($c_{\tilde{W}}$) up to 1 are considered. Effective scales $f_{a}$ are explored in the range 1--\SI{5}{\TeV}.

The generation of signal samples for 2HDM+$a$ is done separately for the gluon--gluon fusion process and the $b$-initiated production. The gluon--gluon fusion process is simulated using the 4-flavour scheme, whereas the $b$-initiated process is calculated using the 5-flavour scheme. The samples are generated using the following parameters: $m_A=m_H=m_{H^{\pm}}$, $\lambda_{P1}=\lambda_{P2}=\lambda_3=3$, $g_{\chi}=1$, and in the alignment limit, $\sin(\beta-\alpha)=1$. The choice of setting $m_H = m_{H^{\pm}}$ is made to evade the constraints from electroweak precision measurements~\cite{Bauer:2017ota}, while the requirement $m_A = m_H$ is introduced to reduce the number of independent model parameters~\cite{2HDMWGproxi}. The nominal samples were produced with $\tan\beta$ of 1 and 10, and $\sin\theta$ of 0.35 and 0.7.

Five benchmark scenarios recommended by the LHC Dark Matter Working Group~\cite{2HDMWGproxi} are used:
\begin{itemize}
\item Benchmark scenario 1: 2D scans in the $m_a$-$m_A$ plane, assuming $\tan \beta = 1.0$ with $\sin \theta = 0.35$ or $\sin \theta = 0.7$,
\item Benchmark scenario 2: 2D scans in the $m_a$-$\tan \beta$ plane, assuming $m_A=\SI{600}{\GeV}$ with $\sin \theta = 0.35$ or $\sin \theta = 0.7$,
\item Benchmark scenario 3: 2D scans in the $m_A$-$\tan \beta$ plane, assuming $m_a=\SI{250}{\GeV}$ with $\sin \theta = 0.35$ or $\sin \theta = 0.7$,
\item Benchmark scenario 4: 1D scans in the $\sin \theta$ range, assuming $\tan \beta = 1.0$ with the fixed masses of $m_A = 600\ (1000)$~GeV and $m_a = 200\ (350)$~GeV, and
\item Benchmark scenario 5: 1D scan in the $m_{\chi}$ range, assuming $m_A=\SI{1000}{\GeV}$, $m_a=\SI{400}{\GeV}$, $\tan \beta = 1.0$, and $\sin \theta = 0.35$.
\end{itemize}

To probe the invisible branching ratio of the Higgs boson, signal samples for all Higgs-boson production modes are generated by forcing the Higgs boson to decay into two $Z$ bosons, which then decay into neutrinos. The Higgs boson mass is set to \SI{125}{\GeV} and the branching ratio to invisible particles is set to 100\%. The cross-sections used for the various production processes follow the Higgs cross-section Yellow Report for $m_h=\SI{125}{\GeV}$~\cite{Heinemeyer:2013tqa}.

The simplified DM model signal samples are generated with vector or axial-vector mediator masses $m_{Z'_{V/A}}$ in the range \SI{100}{\GeV} to \SI{1.7}{\TeV}. The mediator coupling parameters are set to $g_q=0.25$, $g_l=0$ and $g_\chi=1.0$ as recommended by the LHC DM Working Group~\cite{DM_DD}, %
and the width of the mediator is computed as the minimal width allowed given the couplings and masses. In the on-shell regime, the signal acceptance $\times$ efficiency is constant as a function of the DM mass $m_\chi$ for a fixed $m_{Z'_{V/A}}$, and the signal generation is mainly performed for $m_\chi=\SI{1}{\GeV}$ which allows a good interpolation of the exclusion limits for other mass points. Signal samples at the diagonal $m_{Z'_{V/A}}=2m_\chi$ are simulated for each value of $m_{Z'_{V/A}}$ between \SI{100}{\GeV} and \SI{1.2}{\TeV}, and a few additional signal samples are also simulated in the off-shell region of the parameter space with the $m_\chi$ and $m_{Z'_{V/A}}$ chosen to be close to the diagonal.


%

\section{Object reconstruction and identification}
\label{sec:reconstruction}

This section describes the reconstruction and identification of jets and leptons (electrons, muons and $\tau$-leptons) used in the calculation of the missing transverse momentum and the definitions of the signal and control regions. Two types of jets are used to reconstruct the hadronically decaying $V$ boson: small-radius ("small-$R$") and large-radius ("large-$R$") jets. Leptons are vetoed in the signal regions, while electrons and muons are used to define the control regions.

Small-$R$ jets are reconstructed by clustering particle-flow objects~\cite{PERF-2015-09} using the \antikt algorithm~\cite{Cacciari:2008gp,Fastjet} with a radius parameter \(R=0.4\), and are used to identify a $V$ boson with a relatively low boost. The jets are calibrated using a jet-energy scale derived from $\sqrt{s}=\SI{13}{\TeV}$ data and simulation~\cite{JETM-2018-05}. Small-$R$ jets are required to have transverse momentum $\pt > \SI{20}{\GeV}$ and pseudorapidity $|\eta| < 4.5$ when used in the \met calculation and $|\eta| < 2.5$ when used in the signal and control regions. Jets with transverse momentum between 20 and \SI{60}{\GeV} and $|\eta| < 2.5$ must satisfy the requirements of the jet vertex tagger (JVT)~\cite{PERF-2014-03} at the \textit{medium} working point.  This suppresses pile-up jets from other proton--proton interactions in the same and neighbouring bunch crossings. Jets containing $b$-hadrons ($b$-tagged jets) are identified using the multivariate $b$-tagging algorithm (DL1r)~\cite{FTAG-2019-07} with a working point that provides a $b$-tagging efficiency of \SI{77}{\percent} on average in simulated \ttbar events.

Large-$R$ jets are used for the reconstruction of boosted hadronically decaying vector bosons. These jets are built from topological clusters, which are calibrated to the hadronic scale using the local hadronic cell weighting scheme~\cite{PERF-2014-07}. They are reconstructed using the anti-\kt algorithm with a radius parameter $R = 1.0$ and groomed with a trimming procedure~\cite{Krohn:2009th} to
reduce the impact of pile-up.
The groomed jets are then calibrated to the jet energy scale and jet mass scale following the techniques described in Ref.~\cite{JETM-2018-02}. Large-$R$ jets are required to satisfy $\pt > \SI{200}{GeV}$ and $|\eta| < 2.0$. In the $\MET+V$ search, the large-$R$ jets are tagged as originating from a hadronic $V$ boson decay using \pt-dependent requirements on the jet mass, substructure variable $D_2$~\cite{D2}, and number of tracks which are ghost-associated~\cite{Matteo_2008} to the ungroomed jet, at a working point with a \SI{50}{\percent} signal efficiency~\cite{JETM-2018-03,ATL-PHYS-PUB-2020-017}. Variable-radius track jets~\cite{ATL-PHYS-PUB-2014-013} are used to identify large-$R$ jets containing $b$-hadrons and are reconstructed from ID tracks using the anti-\kt algorithm, where only tracks with $\pt > \SI{0.5}{\GeV}$ and a longitudinal impact parameter $|z_0 \sin\theta|<\SI{3}{\mm}$ are used. The variable-radius track jets are required to satisfy $\pt > \SI{10}{GeV}$ and $|\eta| < 2.5$, and are matched to large-$R$ jets using $\Delta R\equiv \sqrt{\Delta{\eta^2}+\Delta{\phi^2}}$. The DL1r discriminant at the 77\% $b$-tagging efficiency working point is also used for this collection.

Electrons are reconstructed by matching tracks found in the ID to clusters of energy deposited in the electromagnetic calorimeter. They must satisfy the \textit{loose} likelihood criteria with $p_\text{T}>\SI[parse-numbers=false]{7}{\GeV}$ and $|\eta|<2.47$~\cite{EGAM-2018-01}, with their energy calibrated as described in Ref.~\cite{EGAM-2021-02}. Additional criteria are applied in the control regions defined using electrons; they must satisfy the \textit{medium} likelihood criteria and have a higher threshold of $p_\text{T}>\SI[parse-numbers=false]{25}{\GeV}$. To suppress jets misidentified as electrons, electrons are required to be isolated as defined by two fixed criteria on track- and calorimeter-based isolation variables~\cite{EGAM-2021-01}. The \textit{tight} and \textit{loose} isolation working points are used for electron selection and veto, respectively. In addition, to suppress electrons not originating from the primary vertex, requirements are set on the longitudinal impact parameter, $|z_0\sin{\theta}|<\SI[parse-numbers=false]{0.5}{\mm}$, and the transverse impact parameter significance, $|d_0|/\sigma(d_0)< 5$.

Muons are reconstructed making use of the information from the ID and the MS~\cite{MUON-2018-03,MUON-2022-01}. All muons must fulfil the \textit{loose} identification criteria with $p_\text{T}>\SI[parse-numbers=false]{7}{\GeV}$ and $|\eta|<2.7$, and satisfy $|z_0\sin{\theta}|<\SI[parse-numbers=false]{0.5}{\mm}$ and $|d_0|/\sigma(d_0)< 3$. In addition, the muons used to define the control regions must satisfy the \textit{medium} identification working point with $p_\text{T}>\SI[parse-numbers=false]{25}{\GeV}$ and $|\eta|<2.5$. All muons are required to be isolated using the \textit{loose} working point. %

Hadronically decaying $\tau$-lepton candidates are reconstructed by combining information from the ID and calorimeters~\cite{ATLAS-CONF-2017-029}. The $\tau$-lepton reconstruction algorithm is seeded by reconstructed small-$R$ jets with $\pt>\SI{10}{GeV}$ and $|\eta|<2.5$, with the reconstructed energies of the $\tau$-lepton candidates corrected to the $\tau$-lepton energy scale~\cite{ATLAS-CONF-2017-029}. $\tau$-lepton candidates must satisfy the \textit{loose} working point, have $\pt > \SI{20}{GeV}$ and $|\eta|<2.5$, excluding the transition region between the electromagnetic barrel and endcap calorimeters ($1.37 < |\eta| < 1.52$), and have one or three associated charged tracks.

The ambiguities among objects are resolved by following an overlap removal procedure that uses the geometrical variable $\Delta{R}$. If two electrons share the same inner-detector track, then only the electron with the higher \pt is considered. An electron is rejected if it shares the same inner-detector track with a muon. If a small-$R$ jet and an electron or a $\tau$-lepton have $\Delta{R} < 0.2$, then the electron or $\tau$-lepton is rejected if the small-$R$ jet is $b$-tagged; otherwise the overlapping jet is removed. Small-$R$ jets are also discarded if they lie within $\Delta{R} < 0.2$ of a muon, are not $b$-tagged and have fewer than three associated tracks with $\pt > \SI{500}{MeV}$. A lepton lying within $\Delta{R} < 0.4$ of a small-$R$ jet that survived all previous overlap criteria is rejected. If a large-$R$ jet and an electron have $\Delta{R} < 1.0$, the overlapping jet is removed.
A variable-radius track jet is rejected if a lepton (electron, muon or $\tau$-lepton) lies within the cone defined by its variable radius.
Hadronically decaying $\tau$-leptons that lie within $\Delta{R} < 0.2$ of an electron or muon are rejected, and any remaining jet within $\Delta{R} < 0.2$ of a hadronically decaying $\tau$-lepton is removed.

The vector missing transverse momentum (\textbf{\met}) is defined as the negative vector sum of the transverse momenta of fully calibrated jets and leptons (electrons, muons and $\tau$-leptons), along with a track soft term~\cite{JETM-2020-03} that includes other tracks reconstructed in the ID. A closely related quantity, the vector missing transverse momentum with invisible leptons (\textbf{\METiell}), is defined in the same way but treats leptons as invisible particles. These two quantities are identical for events without leptons. The \METiell distribution\footnote{The magnitudes of missing transverse momentum variables are denoted as: $\METiell\equiv|\textbf{\METiell}|$.} of $V$+jets processes that contains a leptonic decay can be used to study the \MET distribution of the $V$+jet backgrounds in the signal region. In addition, the vector track-based missing transverse momentum (\textbf{\ptmiss}) and the vector track-based missing transverse momentum with invisible leptons (\textbf{\ptmissiell}) follow the same definitions but use only charged tracks.

The definition of \met significance $\mathcal{S}$ is: \[\mathcal{S}=\frac{\MET}{{[\sigma_L^2(1-\rho_{LT}^2)]}^{1/2}}\] where $\sigma_L^2$ is the total standard deviation in the longitudinal direction (parallel to $\textbf{\met}$), corresponding to the summation of the covariance matrices describing the resolutions of the objects used in the $\textbf{\met}$ calculation, and $\rho_{LT}$ is the correlation factor of the longitudinal $L$ and transverse $T$ measurements. In this way, $\mathcal{S}$ takes into account the expected resolutions of all the objects that enter the \met calculation and their directional correlations~\cite{ATLAS-CONF-2018-038}.


%

\section{Event selection}
\label{sec:selection}
The events considered in this search must satisfy a set of preselection criteria and the requirements of one of the signal regions (SRs) or control regions (CRs).

The selection starts by requiring that the event satisfies one of the unprescaled \met trigger selections~\cite{TRIG-2019-01} with thresholds between 70 and \SI{110}{\GeV}, depending on the data-taking period, or the lowest unprescaled single electron trigger, depending on the region. The \MET calculation at trigger-level does not include muon information, which is why \MET triggers can be used for signatures containing muons, but not those containing electrons. Data cleaning requirements are applied to reject non-collision backgrounds. Events are required to have a primary vertex with at least two associated tracks with $\pt > $\SI{500}{\MeV}. The hard scatter vertex is selected as the one with the largest sum of the square of the transverse momenta of the associated tracks. To suppress multijet events where significant $\met$ originates from jet energy mismeasurements in the calorimeters, the azimuthal angles between \textbf{\METiell} and all small-$R$ jets ($j_i$) or \textbf{\ptmissiell} need to satisfy $\text{min}_i(\Delta\phi(\textbf{\METiell}, j_i))> \ang{20}$ and $\Delta\phi(\textbf{\METiell}, \textbf{\ptmissiell})< \ang{90}$, and \ptmissiell must exceed \SI{30}{\GeV}. %

The merged topology requires $\METiell>\SI{250}{\GeV}$, at least one large-$R$ jet ($J$) and up to four small-$R$ jets. The upper bound on the number of small-$R$ jets is imposed because this number is expected to be small for the signal models probed. The azimuthal angle between \textbf{$\METiell$} and the leading large-$R$ jet, $\Delta \phi(\textbf{\METiell}, J_1)$, is required to be larger than \ang{120}. To suppress background processes with heavy-flavour jets, events with $b$-tagged track jets not associated with the leading large-$R$ jet are vetoed.%
The merged topology event selection is further split into high purity and low purity subsets to enhance the sensitivity, labelled with \texttt{MHP} (merged high purity) and \texttt{MLP} (merged low purity). Events that satisfy all selection criteria from the $V$ boson tagger for the leading large-$R$ jet are classified into the \texttt{MHP} subset. If an event satisfies the requirement on the jet mass but fails the requirements on the substructure variable $D_{2}$ or the number of tracks, it is classified into the \texttt{MLP} subset.

Events that do not satisfy the merged topology selection are considered for the resolved topology selection.  The resolved topology requires $\METiell > \SI{200}{GeV}$ and two to four small-$R$ jets, with the leading jet $\pt$ larger than $\SI{45}{\GeV}$ and the scalar sum of the transverse momenta of all small-$R$ jets larger than 120~(150)$~\GeV$ in events with two (more than two) small-$R$ jets. Additionally, the azimuthal angle between \textbf{$\METiell$} and the leading dijet system is required to satisfy $\Delta \phi(\textbf{\METiell}, j_1j_2)>\ang{120}$. To select a dijet system consistent with a hadronic $V$ boson decay, the angles between the two leading small-$R$ jets ($j_1,j_2$) must satisfy $\Delta \phi (j_1,j_2)<\ang{140} $ and $\Delta R (j_1,j_2)<1.4$, and the invariant mass of the dijet system must be consistent with a $V$ boson, $65 \le m_{j_1j_2} \le \SI{105}{\GeV}$.

In the SRs, a veto is applied on events with electrons, muons or $\tau$-leptons and on events with an \MET significance lower than 8. The signal regions are labelled \texttt{SRMHP} for the merged topology with a high purity, \texttt{SRMLP} for the merged topology with a low purity, and \texttt{SRR} for the resolved topology.

The dominant background contributions, discussed in more details below, are constrained in background-enriched CRs defined using leptons. Four CRs are defined for each SR:
\begin{itemize}
\item[•] di-leptonic CRs are required to have two leptons ($ee$ or $\mu\mu$) to estimate the contribution of the $Z$+jets process, and
\item[•] single-leptonic CRs are required to have one muon. They are split using different $b$-tagged jet multiplicities to distinguish between $W$+jets and $t\bar{t}$ processes.
\end{itemize}

The dominant background in the signal regions is $Z$+jets where the $Z$ boson decays to two neutrinos. This process can be studied using events in which the $Z$ boson decays to muons or electrons. Similarly, irreducible diboson background contributions involving $Z \rightarrow \nu\nu$ decays such as $Z(\nu\nu)Z(q\bar{q})$ and $Z (\nu\nu) W(q'\bar{q})$ can be constrained in the di-leptonic CRs through the $Z (ll) V(qq)$ process where the $V(qq)$ decay products fall into the $W/Z$ mass window.
The di-leptonic CRs select $Z\rightarrow \mu\mu(ee)$ events by requiring exactly two muons (two electrons), an electron (muon) veto and $m_{\mu\mu(ee)}\in[66,116]\,\GeV$ and are labelled as \texttt{CRMHP2mu} (\texttt{CRMHP2el}), \texttt{CRMLP2mu} (\texttt{CRMLP2el}), and \texttt{CRR2mu} (\texttt{CRR2el}).

The modelling of the $W(\rightarrow \mu\nu )$ + jets process is corrected using CRs defined with one muon and zero $b$-tagged track jets inside the large-$R$ jet (merged) or zero $b$-tagged small-$R$ jets (resolved) in the merged (labelled as \texttt{CRMHP1mu0b} and \texttt{CRMLP1mu0b}) and the resolved topologies (labelled as \texttt{CRR1mu0b}). These CRs select $W\rightarrow \mu\nu$ events by requiring exactly one muon, an electron veto and the transverse mass $m_{\mu\nu}^{\text{T}}\in[30,100]\,\GeV$, defined as $m_{\mu\nu}^{\text{T}} = \sqrt{2\pT^{\mu}\MET(1-\cos(\Delta\phi(\mathbf{\boldsymbol p_{\textbf{T}}^{\boldsymbol\mu}},\textbf{\MET})))}$. %
The modelling of the \ttbar and single top-quark backgrounds is corrected using CRs designed with one muon and at least one $b$-tagged track jet inside the large-$R$ jet in the merged topology (labelled as \texttt{CRMHP1mu1b} and \texttt{CRMLP1mu1b}) or at least one $b$-tagged small-$R$ jet in the resolved topology (labelled as \texttt{CRR1mu1b}). They follow the same definition as the previous CRs, except that at least one $b$-tagged track jet (merged) or at least one $b$-tagged small-$R$ jet (resolved) is required. The SR and CR selections are summarized in Table~\ref{tab:summarySelection} for the merged high purity, merged low purity and resolved topologies.

\begin{sidewaystable}[tbp]
\centering
\caption{Summary of the event selection, with small-$R$ jets ($j$), large-$R$ jets ($J$), the number of $b$-tagged track jets ($n_{b\in J}$) inside the leading large-$R$ jet ($J_{1}$), and the number of $b$-tagged small-$R$ jets ($n_b$).}
\label{tab:summarySelection}
\begin{tabular}{l|c|c|c|c|c|c|c|c|c|c}
\toprule
\midrule
& \multicolumn{5}{c|}{\textbf{Merged}}& \multicolumn{5}{c}{\textbf{Resolved}} \\
\midrule
\multirow{6}{*}{Preselection}& \multicolumn{10}{c}{Data cleaning} \\
& \multicolumn{10}{c}{Primary vertex with at least two tracks with $\pt>\SI{500}{\MeV}$} \\
& \multicolumn{10}{c}{No $\tau$-leptons} \\
& \multicolumn{10}{c}{$\ptmissiell > \SI{30}{\GeV}$} \\
& \multicolumn{10}{c}{$\min_i(\Delta\phi(\textbf{\METiell}, j_i)) > \ang{20}$} \\
& \multicolumn{10}{c}{$\Delta\phi(\textbf{\METiell}, \textbf{\ptmissiell})< \ang{90}$} \\
$ \Delta\phi(\textbf{\METiell}, V) $& \multicolumn{5}{c|}{$ \Delta\phi(\textbf{\METiell}, J_1)>\ang{120} $}& \multicolumn{5}{c}{$ \Delta\phi (\textbf{\METiell},j_1j_2)>\ang{120} $} \\
\METiell & \multicolumn{5}{c|}{$>\SI{250}{\GeV}$}& \multicolumn{5}{c}{$>\SI{200}{\GeV}$} \\
\multirow{3}{*}{Jets} & \multicolumn{5}{c|}{$\geq1J;\ \ \leq4j$}& \multicolumn{5}{c}{$\geq2j;\ \ \leq4j$} \\
& \multicolumn{5}{c|}{$ \pt^{J_1} > \SI{200}{\GeV} $}& \multicolumn{5}{c}{$ \pt^{j_1}> \SI{45}{\GeV} $} \\
& \multicolumn{5}{c|}{$b$-tagged track jet veto outside $J_1$}& \multicolumn{5}{c}{$ \sum_{i} \pt^{j_i} \geq $ 120\,(150)\,\unit{\GeV} for $2j$ ($\geq3j$)} \\ [6pt]
\multirow{2}{*}{V-tag} & \multicolumn{5}{c|}{\textbf{High purity:} mass and substructure}& \multicolumn{5}{c}{$ \Delta\phi (j_1,j_2)<\ang{140};\ \ \Delta R (j_1,j_2)<1.4 $} \\
& \multicolumn{5}{c|}{\textbf{Low purity:} mass and inverted substructure}& \multicolumn{5}{c}{$\twomass{j_1}{j_2} \in [65,105]\,\unit{\GeV}$} \\
\midrule
&  \texttt{SR}  & \texttt{CR2mu} & \texttt{CR2el} & \texttt{CR1mu0b} & \texttt{CR1mu1b} &  \texttt{SR}  & \texttt{CR2mu} & \texttt{CR2el} & \texttt{CR1mu0b} & \texttt{CR1mu1b} \\ \midrule
Trigger        &  \MET   & \MET   & Electron   & \MET     & \MET    &  \MET  & \MET   & Electron   & \MET     & \MET     \\
$e$          &  0   & 0   & 2   & 0     & 0     &  0   & 0   & 2   & 0     & 0     \\
$\mu$          &  0   & 2   & 0   & 1     & 1     &  0   & 2   & 0   & 1     & 1     \\
$\mathcal{S}$      & $>\!8\,\phantom{>}$ & -   & -   & -     & -     & $>\!8\,\phantom{>}$ & -   & -   & -     & -     \\
$\twomass{\ell}{\ell}\,[\unit{\GeV}]$  & - & $\in[66,116]$ & $\in[66,116]$ & - & - & - & $\in[66,116]$ & $\in[66,116]$ & - & - \\
$m_{\mu\nu}^{\text{T}}\,[\unit{\GeV}]$ & - & - & - & $\in[30,100]$ & $\in[30,100]$ & - & - & - & $\in[30,100]$ & $\in[30,100]$ \\
$n_{b\in J}$       & -    & -   & -   & 0     &$\geq\!1\,\phantom{\geq}$& -    & -   & -   & -     & -      \\
$n_{b}$         & -    & -   & -   & -     & -     & -    & -   & -   & 0     & $\geq\!1\,\phantom{\geq}$\\
\midrule
\bottomrule
\end{tabular}
\end{sidewaystable}

A set of criteria to suppress the contribution of multijet events is included as mentioned earlier. Due to its low acceptance in the SRs and the
challenge of accurately modelling non-Gaussian effects in jet reconstruction and resolution, the residual multijet background in the SRs\footnote{The multijet background in the CRs is negligible.} is estimated using a data-driven approach. A template region enriched in multijets is built by inverting the preselection requirement $\min_i(\Delta\phi(\textbf{\METiell}, j_i))<\ang{20}$. A multijet template of the \met spectrum for the range $150 < \met < \SI{600}{GeV}$ is extracted from this sample by subtracting the contribution from other sources of background, as estimated using MC, from data. The normalisation is determined through a fit that uses the $V$ mass upper-side-band region $110 < m_{J/j_1j_2} < \SI{250}{GeV}$ without the $\Delta\phi(\textbf{\met},\textbf{\ptmiss})<\ang{90}$ requirement. The multijet background estimate is small compared with the total MC-based background expectation, and is estimated to be \SI{0.1}{\percent} in \texttt{SRMHP}, \SI{0.2}{\percent} in \texttt{SRMLP} and \SI{0.5}{\percent} in \texttt{SRR}.

%
%


%

\section{Statistical analysis, systematic uncertainties and the background-only model results}
\label{sec:result}

A binned maximum likelihood fit is used for the background estimation and the interpretation of the results. The fit is performed simultaneously on the \MET\ distributions in all SRs and on the \METiell distributions in all CRs.  The fit variable is referred to below as \METiell in all regions for simplicity as the two variables are identical in the SRs. The likelihood is built by multiplying the likelihoods obtained in each region:
\[L(\mu,\boldsymbol{\kappa}, \boldsymbol{\theta})=\prod_{r}\prod_{i}{\mathrm{Poisson}}\big(N_{r\!,i}^{\mathrm{obs}}\, |\, N_{r\!,i}^{\mathrm{sig}}(\mu,\boldsymbol{\theta})+N_{r\!,i}^{\mathrm{bkg}}(\boldsymbol{\kappa}, \boldsymbol{\theta})\big) f_{\mathrm{constr}}(\boldsymbol{\theta})\]
where $i$ is the $i$-th \METiell bin considered in the fit, $r$ denotes the regions (SRs and CRs), $N_{r\!,i}^{\mathrm{obs}}$ is the observed yield in the $i$-th \METiell bin of region $r$, $N_{r\!,i}^{\mathrm{sig(bkg)}}$ is the expected yield of the signal (background) in the $i$-th \METiell bin of the region $r$, $\boldsymbol{\theta}$ is the vector of nuisance parameters including systematic uncertainties in the predicted yield in each region and \METiell bin, $\mu$ is the scale factor associated with the normalisation of the signal model (signal strength), and $\boldsymbol{\kappa}$ is the vector of the normalisation factors of the main backgrounds ($Z$+jets, $W$+jets and $t\bar{t}$). The term $f_{\mathrm{constr}}$ represents the product of the Gaussian constraints applied to each of the nuisance parameters. The unconstrainted parameters in the fit are the signal strength $\mu$ and the normalisation factors $\boldsymbol{\kappa}$. Separate normalisation factors are assigned to the $Z$+jets, $W$+jets and $t\bar{t}$ background normalisations for all three topologies (resolved, merged high purity and merged low purity), resulting in a total of ten normalisation factors, including the signal strength $\mu$. The normalisations of diboson and other electroweak backgrounds are taken from the nominal MC predictions and each is allowed to vary within an assigned uncertainty of \SI{25}{\percent}~\cite{HIGG-2016-29}. The multijet contributions in the SRs are estimated using a data-driven approach, as described in Section~\ref{sec:selection}.

The uncertainites related to the finite number of simulated events are included using Gaussian constraints in each bin where they are larger than \SI{0.5}{\percent} of the total number of simulated events in the bin.

As detailed below, systematic uncertainties in the overall normalisation and the \METiell shape of the signal and background processes are estimated and implemented in the fit. The systematic uncertainties are correlated across regions and background processes, except the process-specific uncertainties, which are only correlated across regions.

Experimental uncertainties include the energy scale and resolution of small-$R$ jets~\cite{JETM-2018-05}, large-$R$ jets~\cite{JETM-2018-02}, electrons~\cite{EGAM-2021-02}, muons~\cite{MUON-2022-01} and $\tau$-leptons~\cite{ATLAS-CONF-2017-029}, as well as reconstruction, identification and isolation uncertainties for electrons~\cite{EGAM-2021-01} and muons~\cite{MUON-2018-03}, and pile-up uncertainties for small-$R$ jets~\cite{ATLAS-CONF-2014-018}. Additional uncertainties regarding the jet mass scale~\cite{ATLAS-CONF-2020-022} and $V$-tagging~\cite{JETM-2018-03,ATL-PHYS-PUB-2020-017} of large-$R$ jets are taken into account, as are $b$-tagging~\cite{FTAG-2018-01,FTAG-2019-02,FTAG-2020-08} and \MET soft term~\cite{JETM-2020-03} uncertainties.

Several theoretical modelling uncertainties are considered for the background processes, and affect mostly the shape of the \METiell distribution. The uncertainties include the effects from varying QCD renormalisation and factorisation scales, the choice of the PDF set, the variation of the strong coupling constant and the modelling of the parton showers. For $V$+jets, additional uncertainties from higher-order electroweak corrections and matching scale variation are considered. For $t\bar{t}$, additional uncertainties arise from the A14 tune, hard scatter generation and parton shower matching. For the $Wt$ background, an uncertainty in the scheme to remove overlaps with the $t\bar{t}$ background is included, where \emph{diagram removal} is used for the nominal fit and the difference with \emph{diagram subtraction}~\cite{Frixione:2008yi} is symmetrised and used as the uncertainty. Theoretical uncertainties in the signal yields due to variations of the renormalisation and factorisation scales, uncertainties in the PDFs along with variations of the strong coupling constant, different generator tunes and the modelling of the parton showers, are estimated for all signal models. In addition, the uncertainty in the initial and final state radiation parameters of the parton shower is considered for the invisible decay of the Higgs boson. An uncertainty of \SI{100}{\percent} in the normalisation of the multijet background is assigned to cover the statistical uncertainty in data, the impact of non-multijet background and the extrapolation from the multijet CR to the SRs.

The impact of a given group of systematic uncertainties on the post-fit signal strength uncertainty is summarised in Table~\ref{syt-tab}. The procedure is done for the invisible Higgs model and one representative point for each of the ALPs, \tHDMa and simplified DM models together, using Asimov datasets which are constructed as the sum of the background expectations and the respective signal model with a signal strength of $\mu=1$. The dominant uncertainty groups are those related to large-$R$ jets (which includes the $V$-tagging uncertainties), muons, $V$+jets modelling and normalisation (dominated by the diboson normalisation uncertainty). The importance of the uncertainty groups varies for different signal models, depending on the \MET shape of the signal. For example, signals with a harder \MET distribution are more sensitive to $V$+jets modelling uncertainties and to data statistical uncertainty, since these increase with \MET. Lepton uncertainties can also have a large impact as they affect the normalisation factors extracted in the CRs and extrapolated to the SRs.

\begin{table}[tbp]
\centering
\caption{Impact of groups of uncertainties on the post-fit signal strength from a fit to the Asimov dataset with $\mu=1$. The impact is defined as the square root of the difference between the squares of the total uncertainty and the uncertainty obtained by neglecting the group of systematic uncertainties in question. It is shown for the signal models (a) ALPs with $m_{a}=\SI{1}{\MeV}$, $c_{\tilde{W}} = 0.2$, and $f_{a} = \SI{3}{\TeV}$, (b) \tHDMa with $\tan\beta=1$, $\sin\theta=0.35$, $m_{A}=\SI{1}{\TeV}$, and $m_{a}=\SI{100}{\GeV}$, (c) the invisible decay of the Higgs boson with $\mathcal{B}_{h\rightarrow inv}$=\SI{100}{\percent}, and (d) the simplified DM axial-vector model with $m_{\chi} = \SI{1}{\GeV}$ and $m_{Z'_{A}} = \SI{1}{\TeV}$.}
\begin{tabular}{l
S[table-format = <2.0, table-column-width = 2.2cm, table-align-comparator = false]
S[table-format = <2.0, table-column-width = 2.2cm, table-align-comparator = false]
S[table-format = <2.0, table-column-width = 2.2cm, table-align-comparator = false]
S[table-format = <2.0, table-column-width = 2.2cm, table-align-comparator = false]
}
\toprule
\midrule
Group of uncertainties & \multicolumn{4}{c}{Impact on $\mu$ uncertainty $\times100$} \\
& {(a)} & {(b)} & {(c)} & {(d)} \\
& {ALP} & {2HDM+$a$}& {$h\rightarrow inv$} & {Simplified DM}\\
& {$c_{\tilde{W}} = 0.2,$} & {$m_{A}=\SI{1}{\TeV}$}  & {$\mathcal{B}_{h\rightarrow inv}$=\SI{100}{\percent}} & {$m_{\chi} = \SI{1}{\GeV},$}  \\
& {$f_{a} = \SI{3}{\TeV}$}  & {$m_{a}=\SI{100}{\GeV}$} &   & {$m_{Z'_{A}} = \SI{1}{\TeV}$} \\ \midrule
Large-$R$ jets           & 24     &  13    & 4 & 13 \\
Small-$R$ jets          & 19     &  6   & 6 & 10 \\
$b$-tagging            & 2      &  1   & 2 & 1 \\
\met                  & 2      &  1   & 4 & 1 \\
Pile-up                & 5      &  2   & 3 & 3 \\
Electron              & 11     &  4   & 7 & 8 \\
Muon                  & 22     &  7  & 9 & 15  \\
$\tau$-lepton          & 1      &  <1  & <1       & <1    \\
Luminosity            & 2      &  1   & 1 & 1  \\
Normalisation (including $\boldsymbol{\kappa}$)  & 23     &  7  & 7 & 15  \\ [6pt]
$V$+jets modelling       & 32     &  11 & 7 & 18  \\
Diboson modelling     & 12     &  4   & 2 & 7 \\
\ttbar modelling       & 5      &  2  & 1 & 2  \\
Multijet estimate    & 1      &  1  & 1 & 1  \\
Single-top modelling   & 1      &  1  & 1 & 1  \\ [6pt]
Signal modelling       & 12     &  1  & 12 & 8  \\ [6pt]
Simulated events sample size & 23     &  11  & 4 & 11 \\ \midrule
Statistical            & 56     &  20  & 6 & 23 \\
Systematic            & 66     &  26  & 19 &35 \\ \midrule
Total                  & 86     &  33  & 20 & 43 \\
\midrule
\bottomrule
\end{tabular}
\label{syt-tab}
\end{table}

A background-only fit ($\mu$ fixed to zero) including all SRs and CRs is performed and results in the \METiell distributions for data and the expected SM backgrounds shown in Figures~\ref{fig:crmhp} -- \ref{fig:bkg_yield}.

\begin{figure}[tbp]
\centering
\begin{subfigure}[t]{0.4\textwidth}
\centering
\includegraphics[width = 0.99\textwidth]{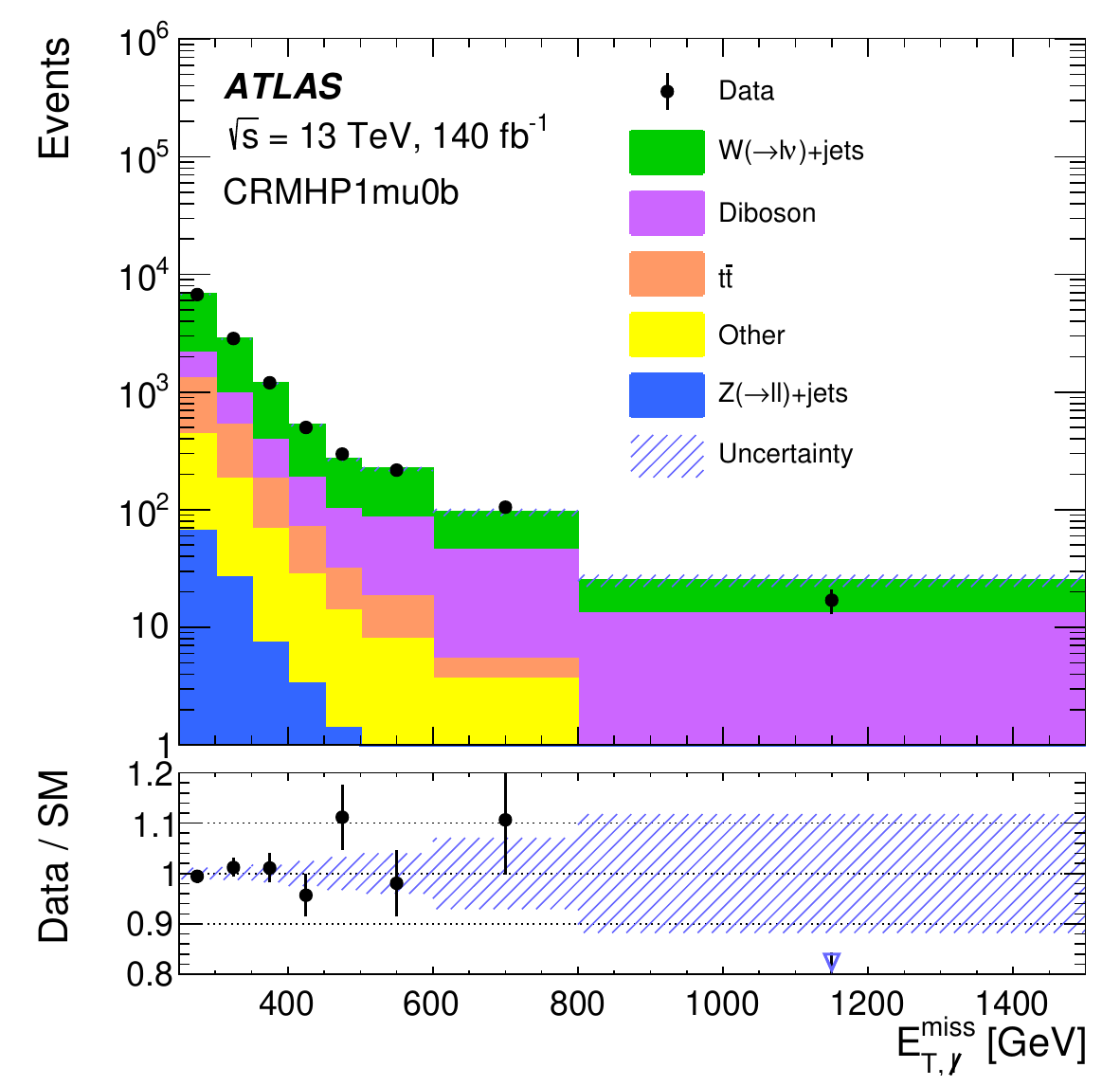}
\caption{}
\end{subfigure}
\begin{subfigure}[t]{0.4\textwidth}
\centering
\includegraphics[width = 0.99\textwidth]{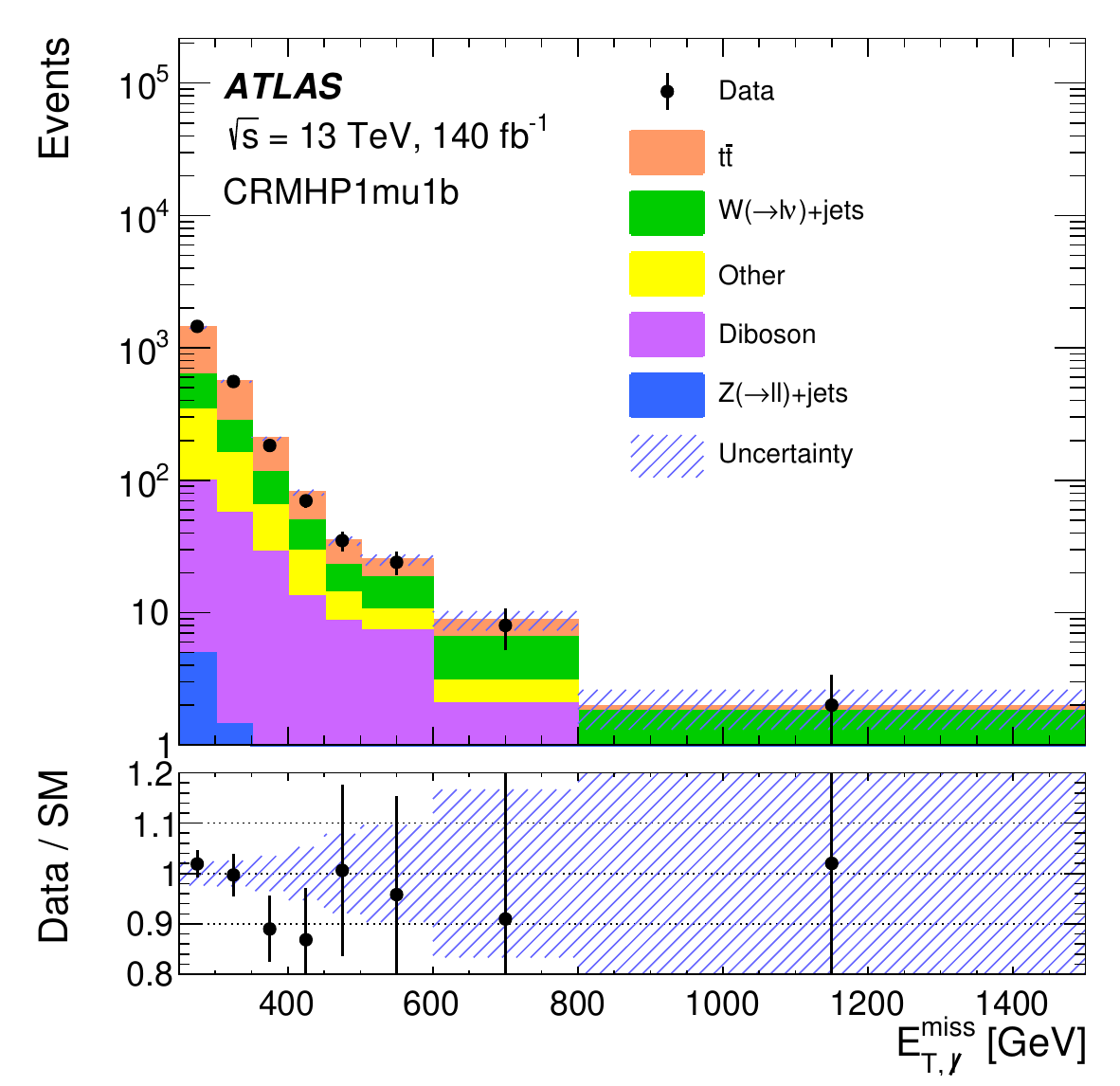}
\caption{}
\end{subfigure}
\begin{subfigure}[t]{0.4\textwidth}
\centering
\includegraphics[width = 0.99\textwidth]{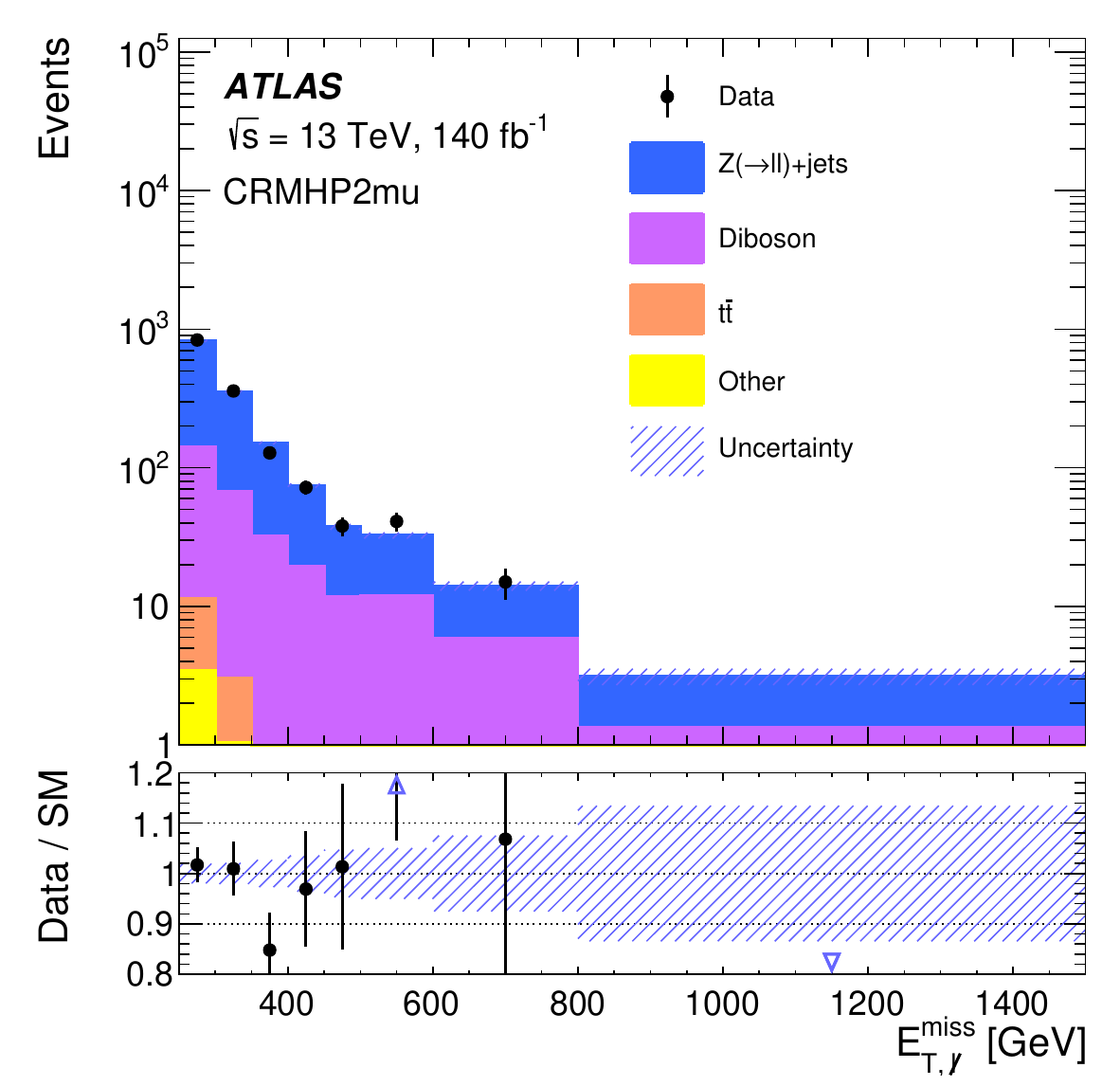}
\caption{}
\end{subfigure}
\begin{subfigure}[t]{0.4\textwidth}
\centering
\includegraphics[width = 0.99\textwidth]{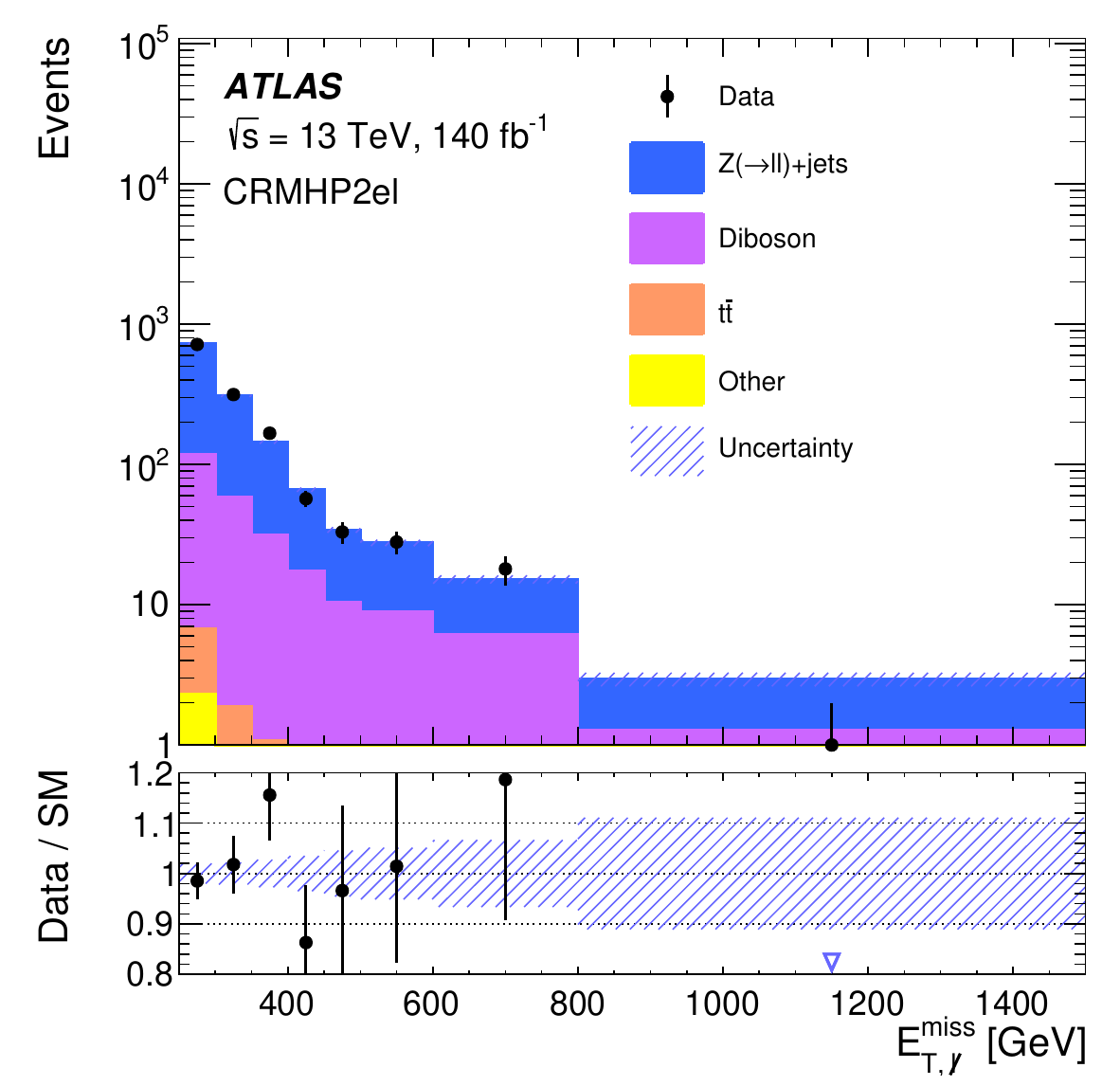}
\caption{}
\end{subfigure}
\caption{Distribution of \METiell after the background-only fit, in the control regions: (a) \texttt{CRMHP1mu0b}, (b) \texttt{CRMHP1mu1b}, (c) \texttt{CRMHP2mu}, and (d) \texttt{CRMHP2el}. The total uncertainty in the background is shown as a shaded band and the error bars on the data points represent the statistical uncertainty. Small background contributions such as $t\bar{t}V$, $Vh$, single-top and triboson processes are grouped into one category and shown as ``Other''. The last bin contains the overflow.}
\label{fig:crmhp}
\end{figure}

\begin{figure}[tbp]
\centering
\begin{subfigure}[t]{0.4\textwidth}
\centering
\includegraphics[width = 0.99\textwidth]{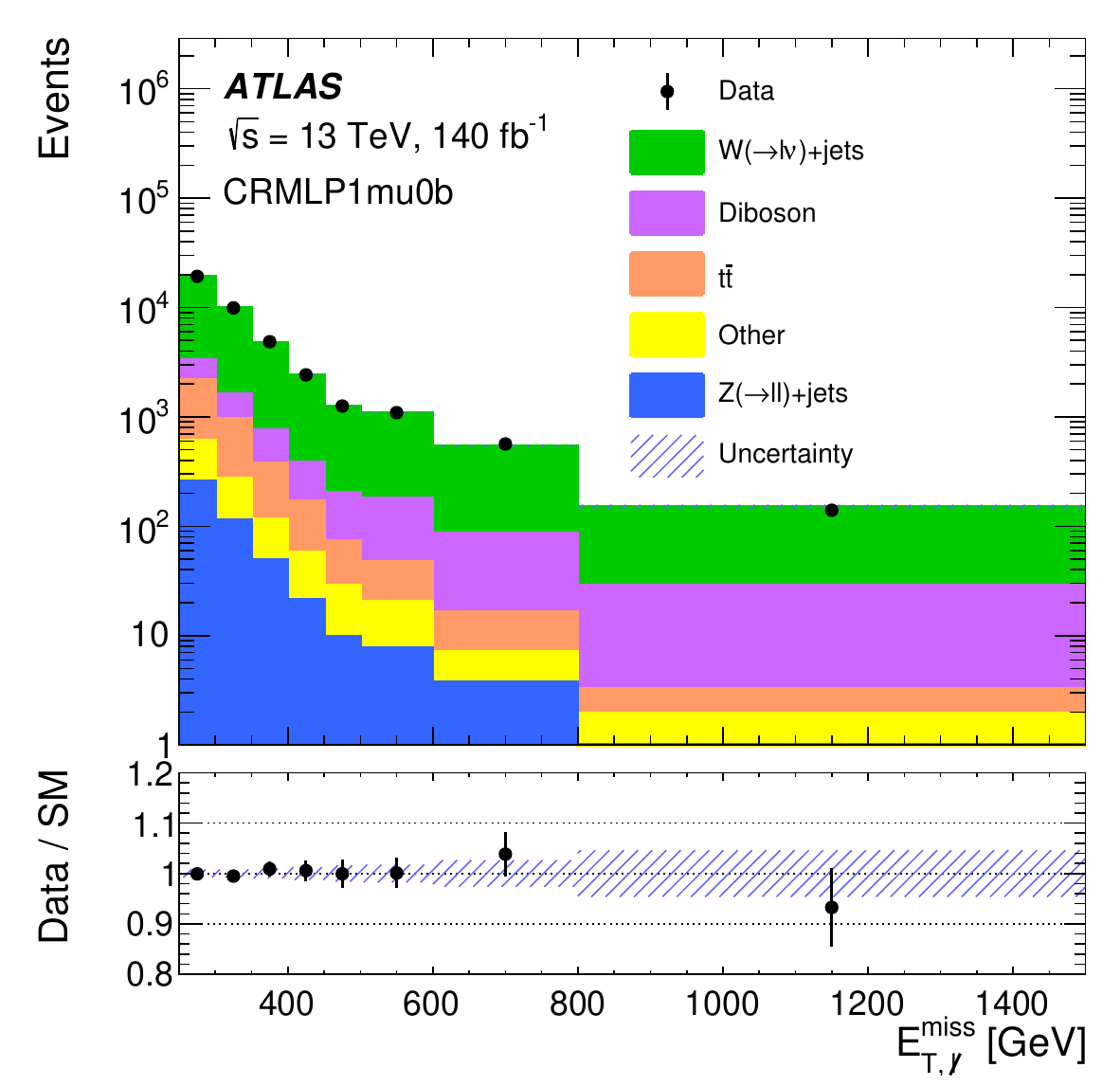}
\caption{}
\end{subfigure}
\begin{subfigure}[t]{0.4\textwidth}
\centering
\includegraphics[width = 0.99\textwidth]{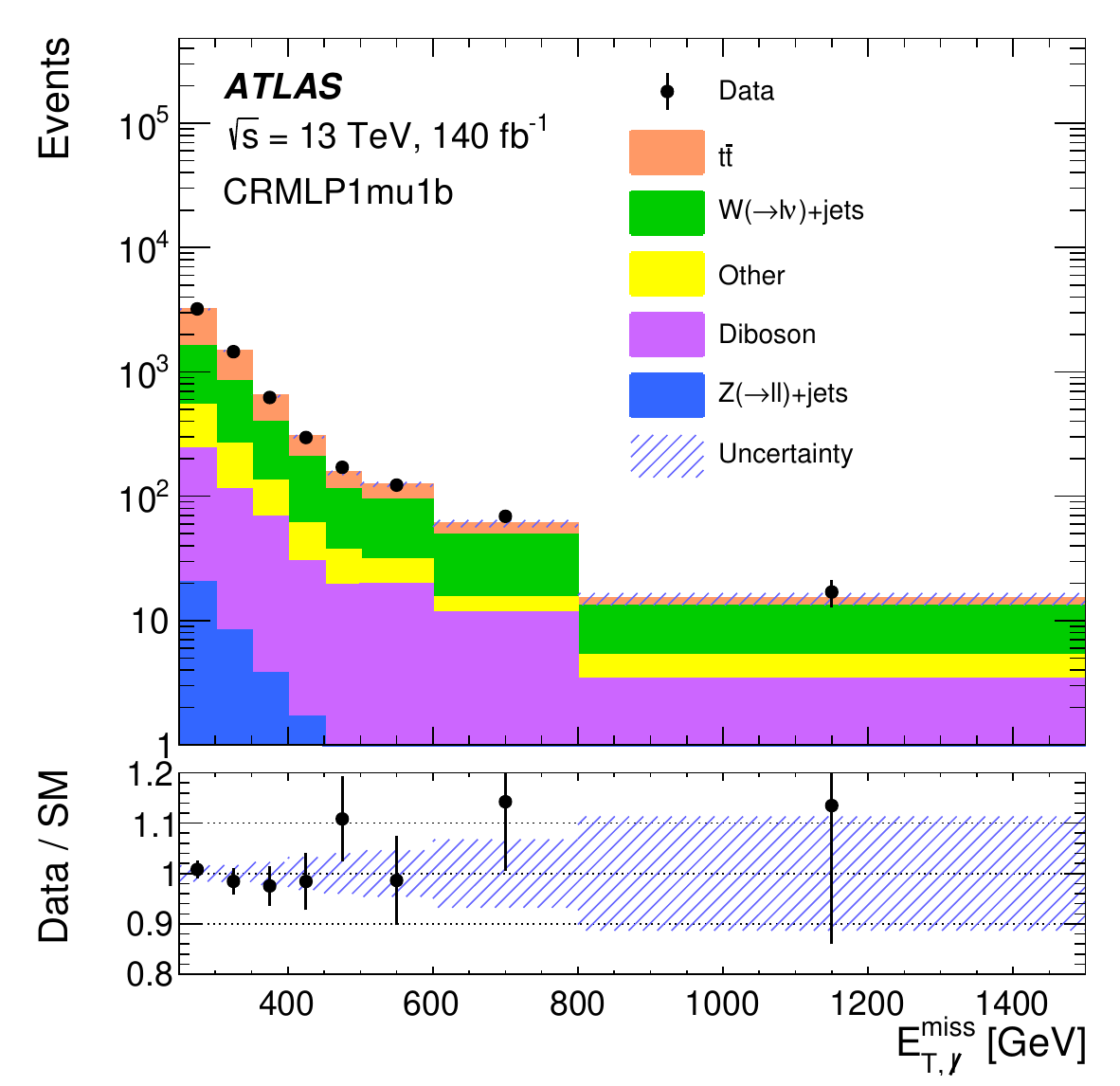}
\caption{}
\end{subfigure}
\begin{subfigure}[t]{0.4\textwidth}
\centering
\includegraphics[width = 0.99\textwidth]{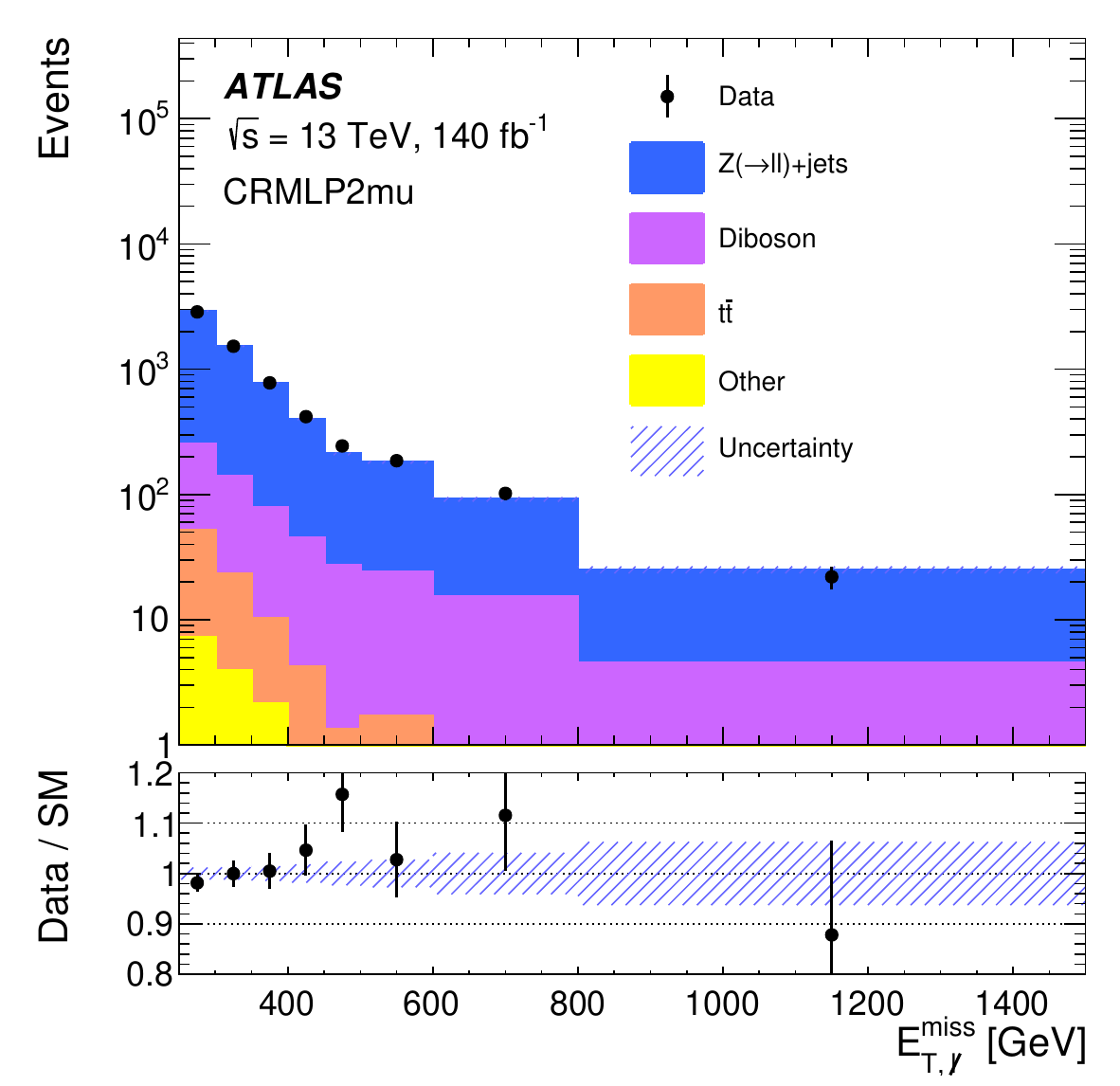}
\caption{}
\end{subfigure}
\begin{subfigure}[t]{0.4\textwidth}
\centering
\includegraphics[width = 0.99\textwidth]{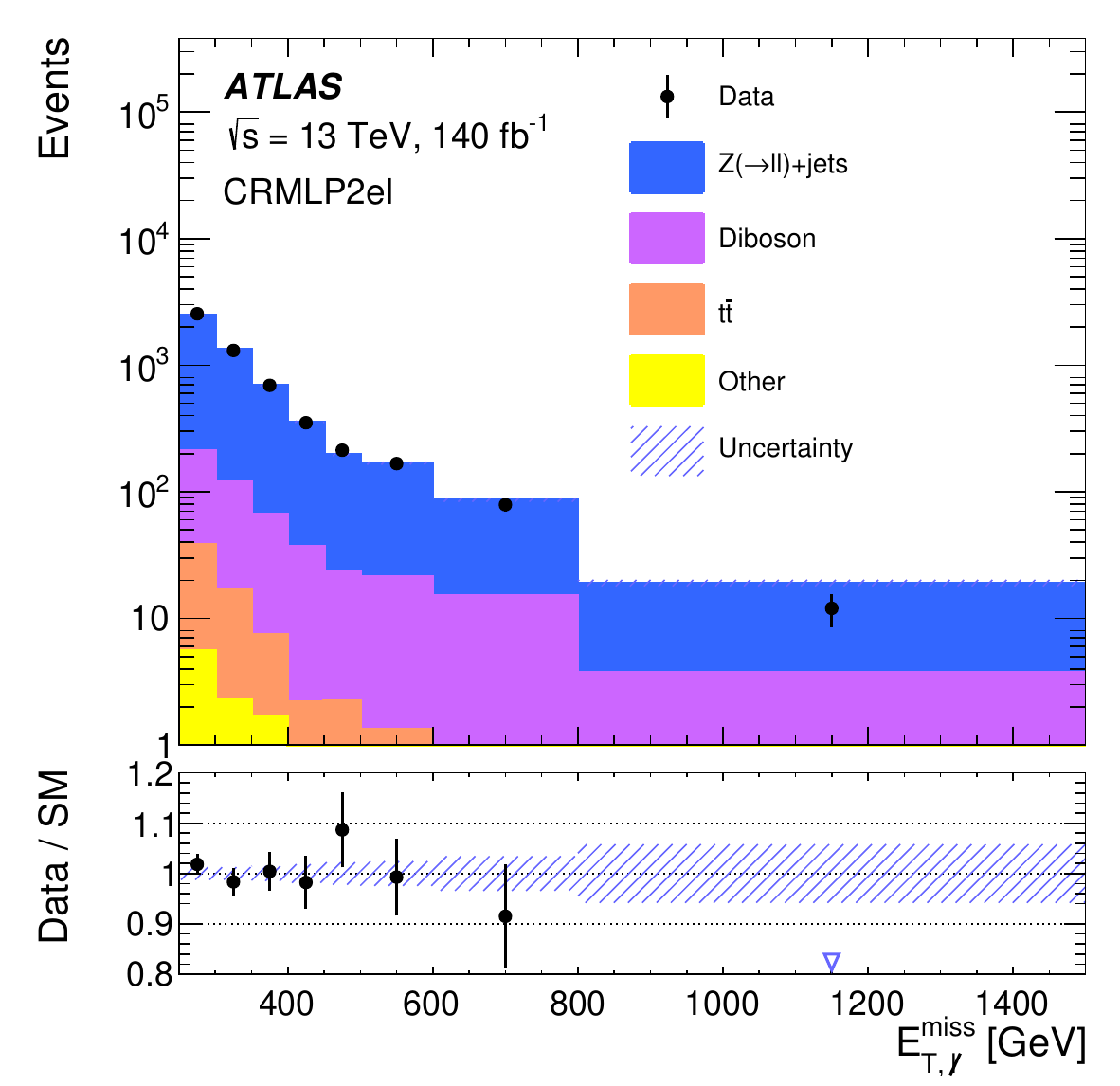}
\caption{}
\end{subfigure}
\caption{Distribution of \METiell after the background-only fit, in the control regions: (a) \texttt{CRMLP1mu0b}, (b) \texttt{CRMLP1mu1b}, (c) \texttt{CRMLP2mu}, and (d) \texttt{CRMLP2el}. The total uncertainty in the background is shown as a shaded band and the error bars on the data points represent the statistical uncertainty. Small background contributions such as $t\bar{t}V$, $Vh$, single-top and triboson processes are grouped into one category and shown as ``Other''. The last bin contains the overflow.}
\end{figure}

\begin{figure}[tbp]
\centering
\begin{subfigure}[t]{0.4\textwidth}
\centering
\includegraphics[width = 0.99\textwidth]{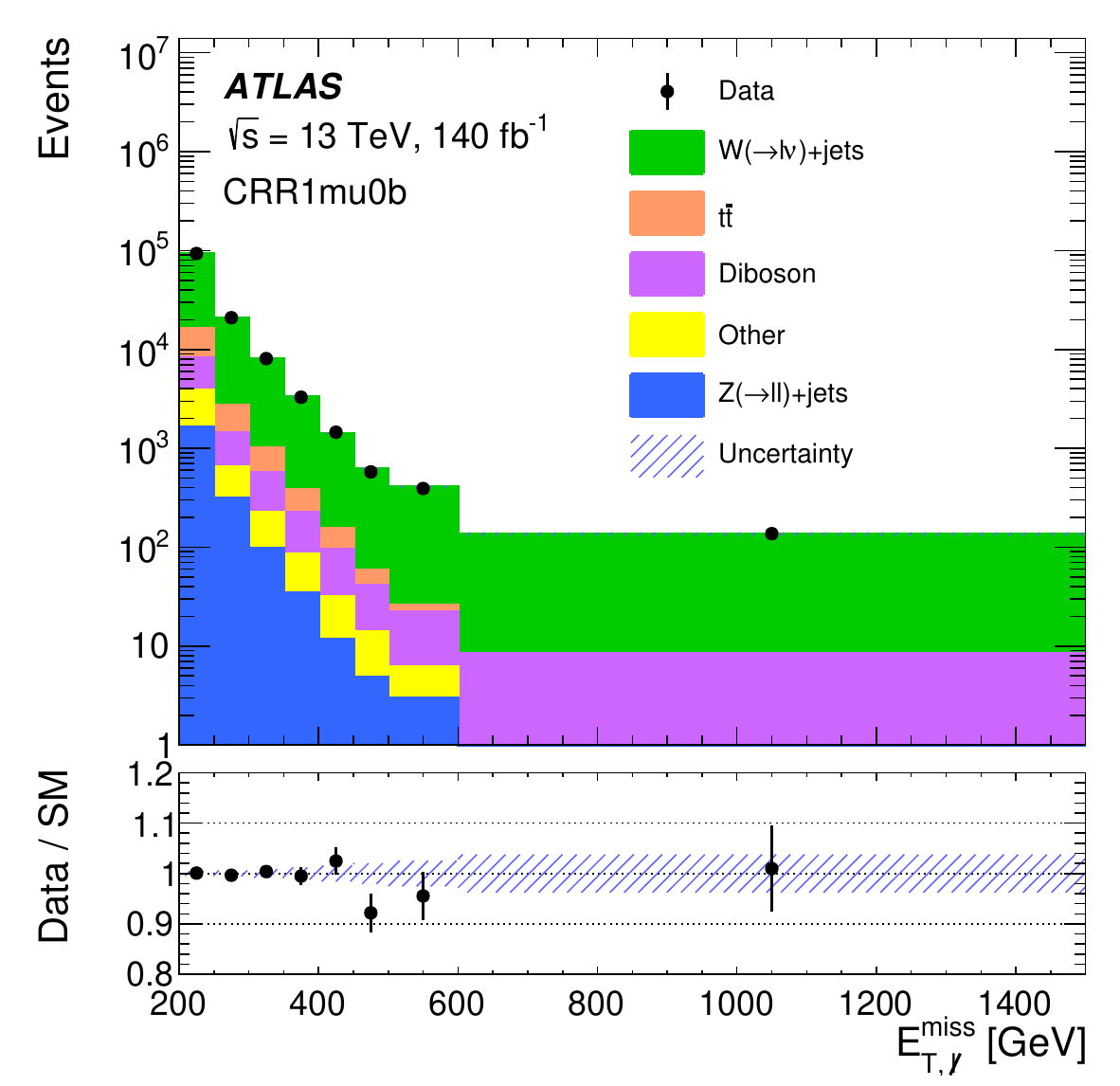}
\caption{}
\end{subfigure}
\begin{subfigure}[t]{0.4\textwidth}
\centering
\includegraphics[width = 0.99\textwidth]{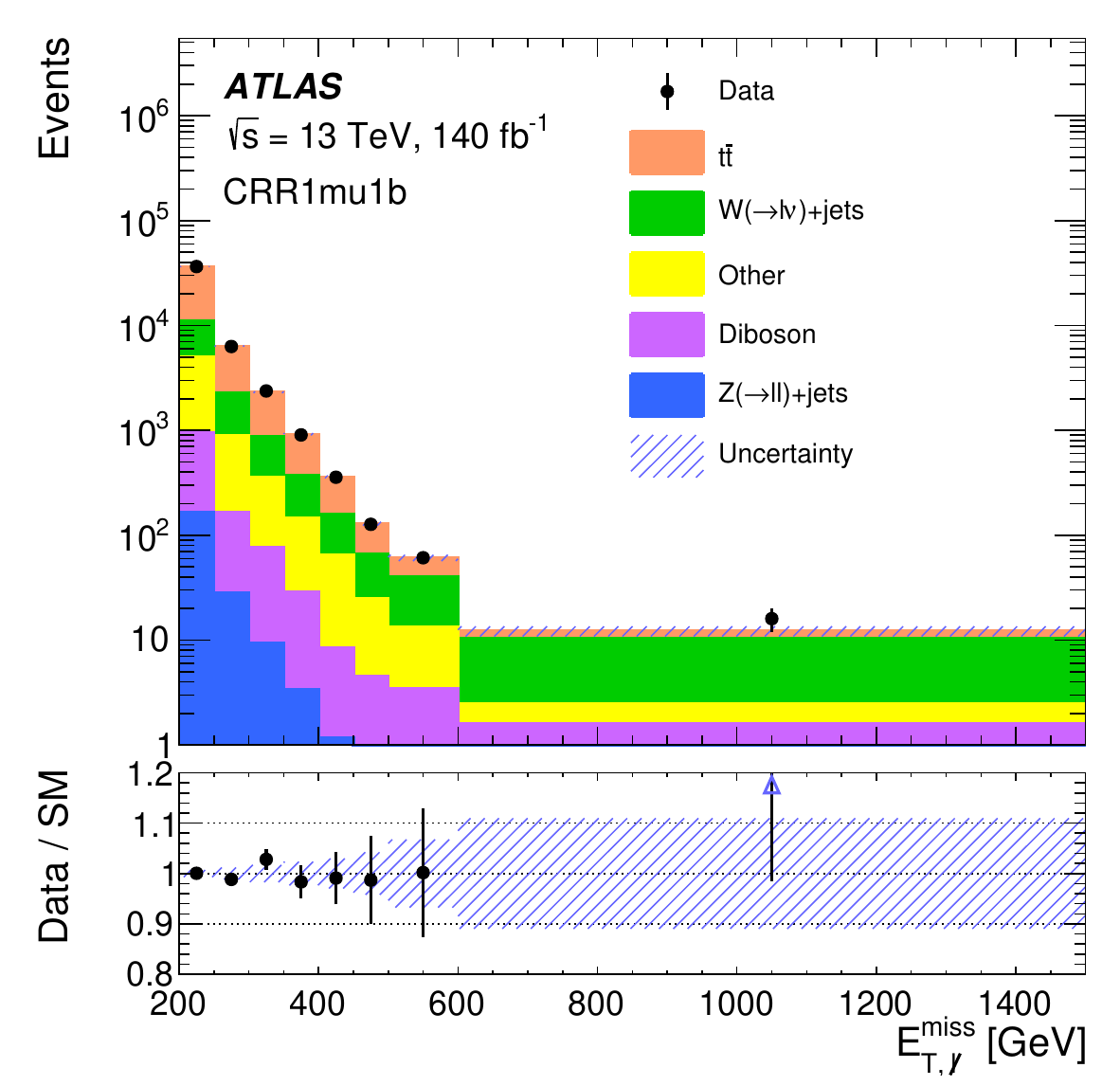}
\caption{}
\end{subfigure}
\begin{subfigure}[t]{0.4\textwidth}
\centering
\includegraphics[width = 0.99\textwidth]{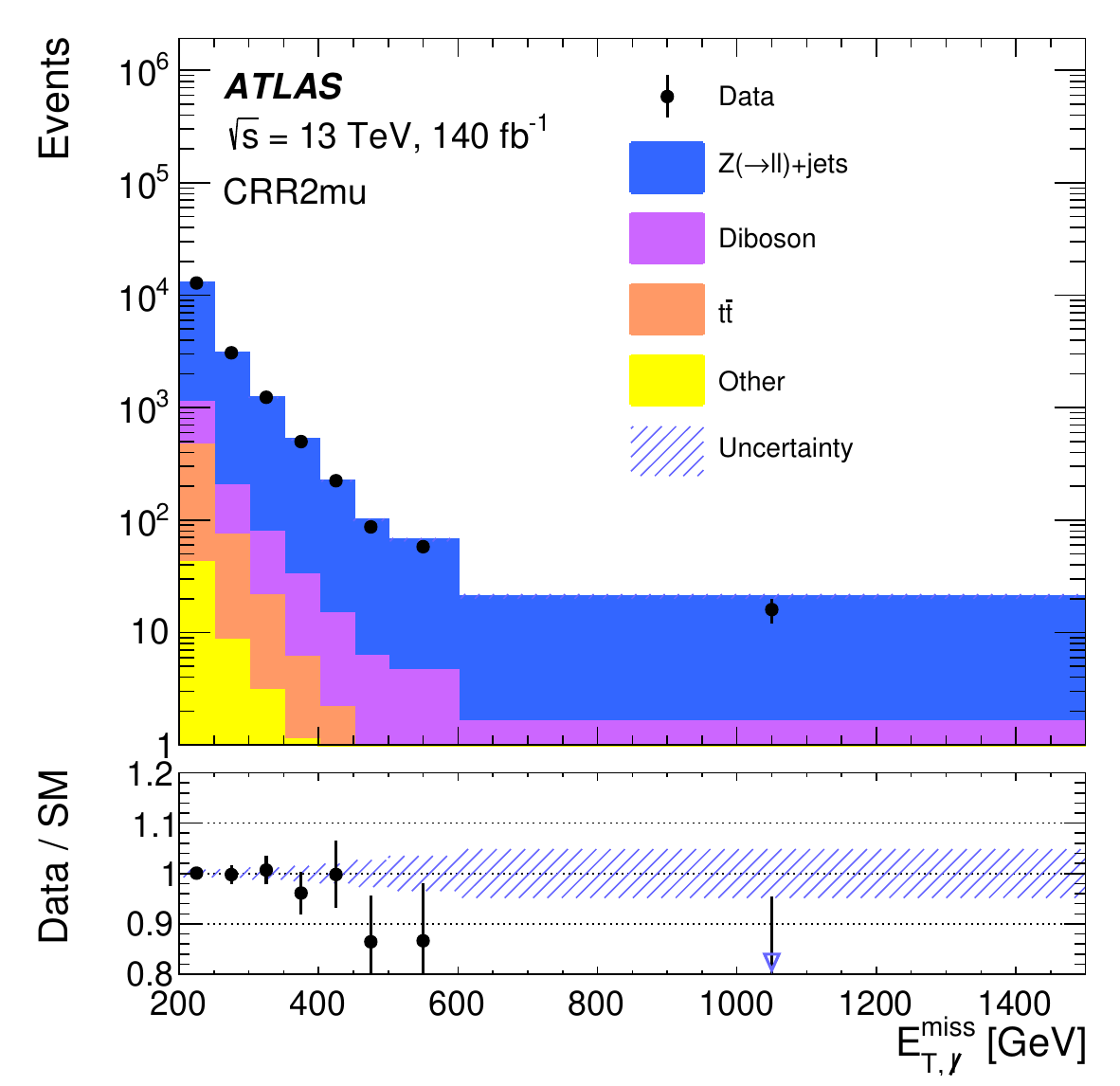}
\caption{}
\end{subfigure}
\begin{subfigure}[t]{0.4\textwidth}
\centering
\includegraphics[width = 0.99\textwidth]{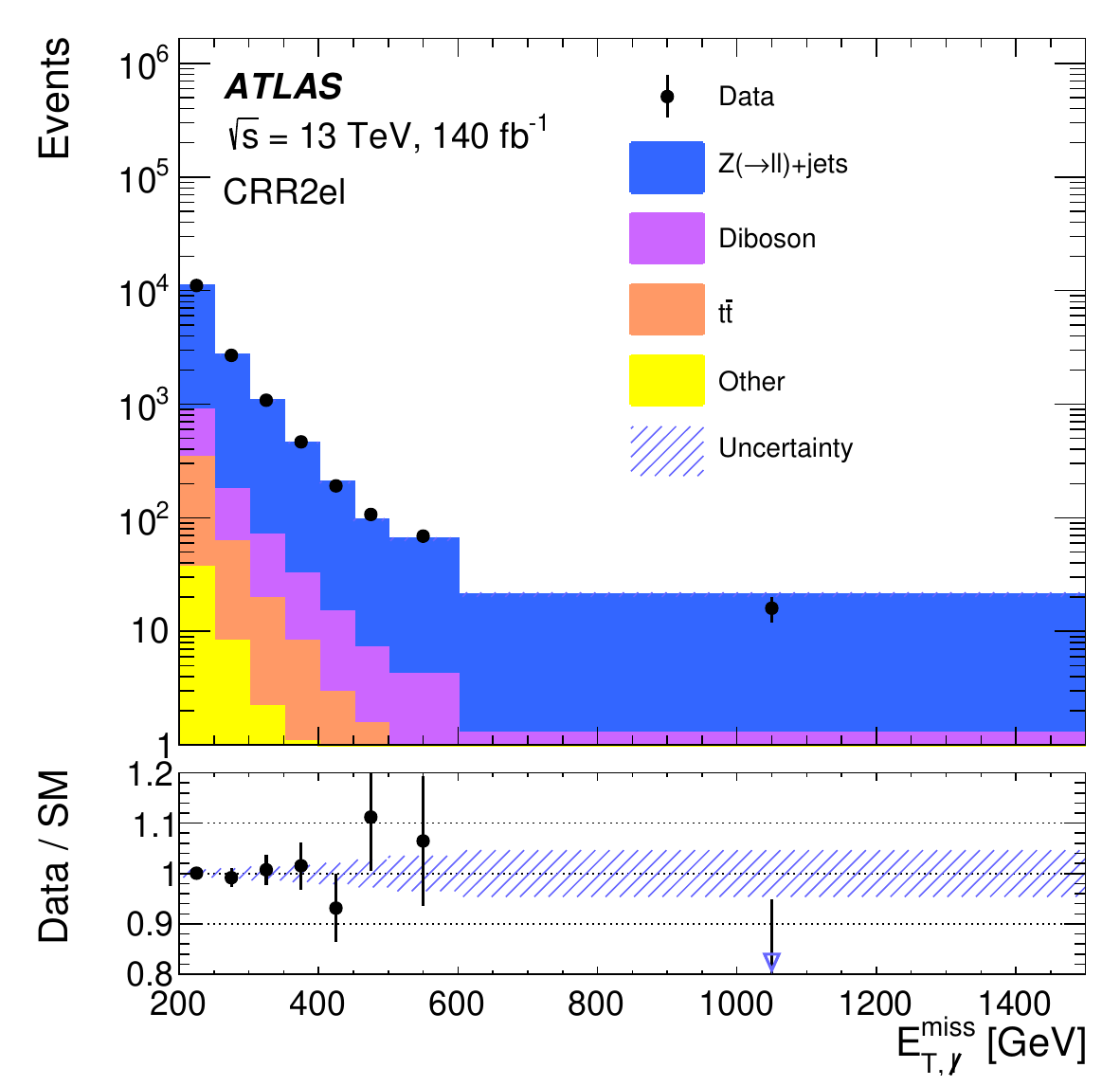}
\caption{}
\end{subfigure}
\caption{Distribution of \METiell after the background-only fit, in the control regions: (a) \texttt{CRR1mu0b}, (b) \texttt{CRR1mu1b}, (c) \texttt{CRR2mu}, and (d) \texttt{CRR2el}. The total uncertainty in the background is shown as a shaded band and the error bars on the data points represent the statistical uncertainty. Small background contributions such as $t\bar{t}V$, $Vh$, single-top and triboson processes are grouped into one category and shown as "Other". The last bin contains the overflow.}
\end{figure}

\begin{figure}[tbp]
\centering
\begin{subfigure}[t]{0.4\textwidth}
\includegraphics[width=0.99\textwidth]{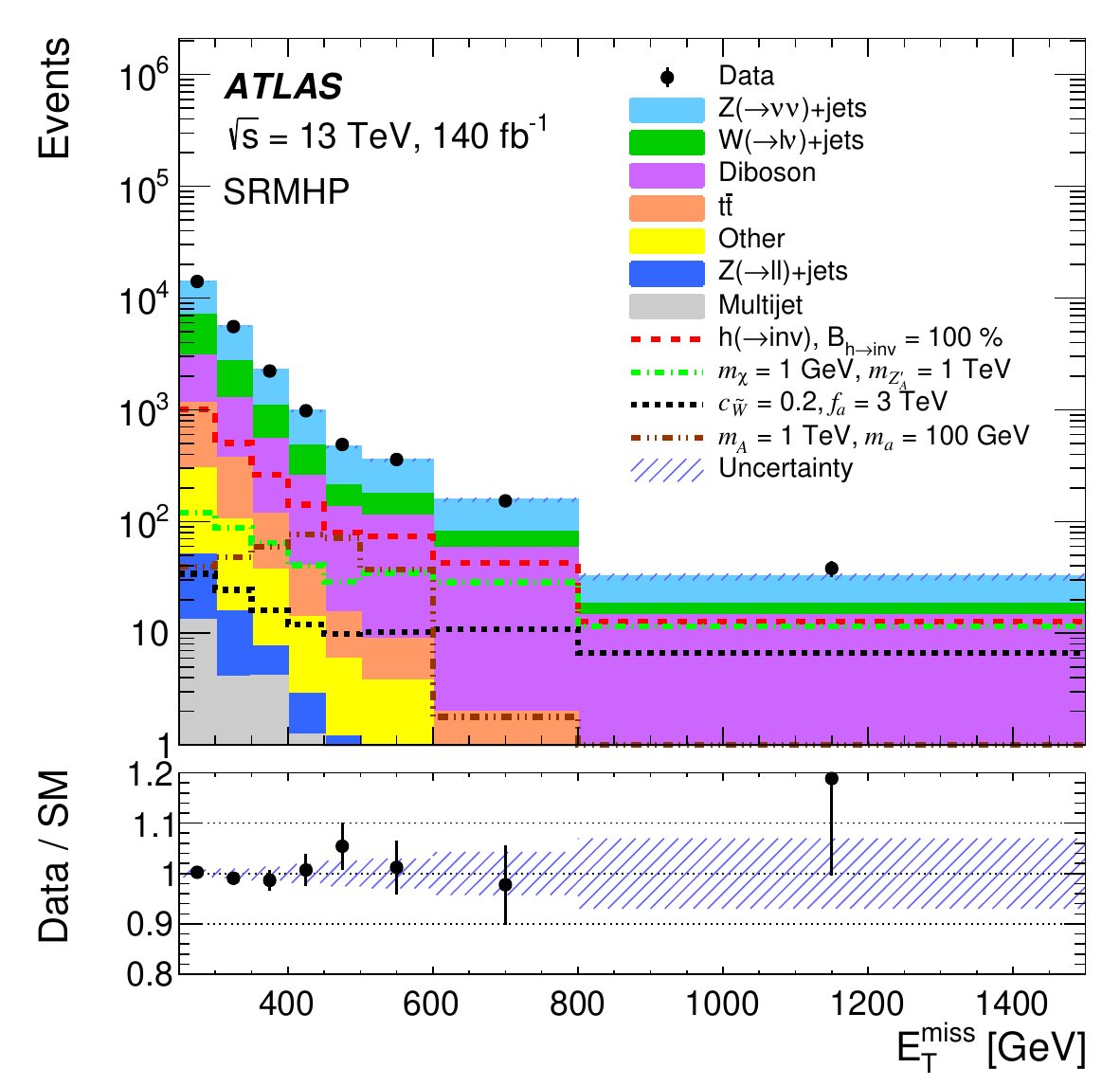}
\caption{}
\end{subfigure}
\begin{subfigure}[t]{0.4\textwidth}
\includegraphics[width=0.99\textwidth]{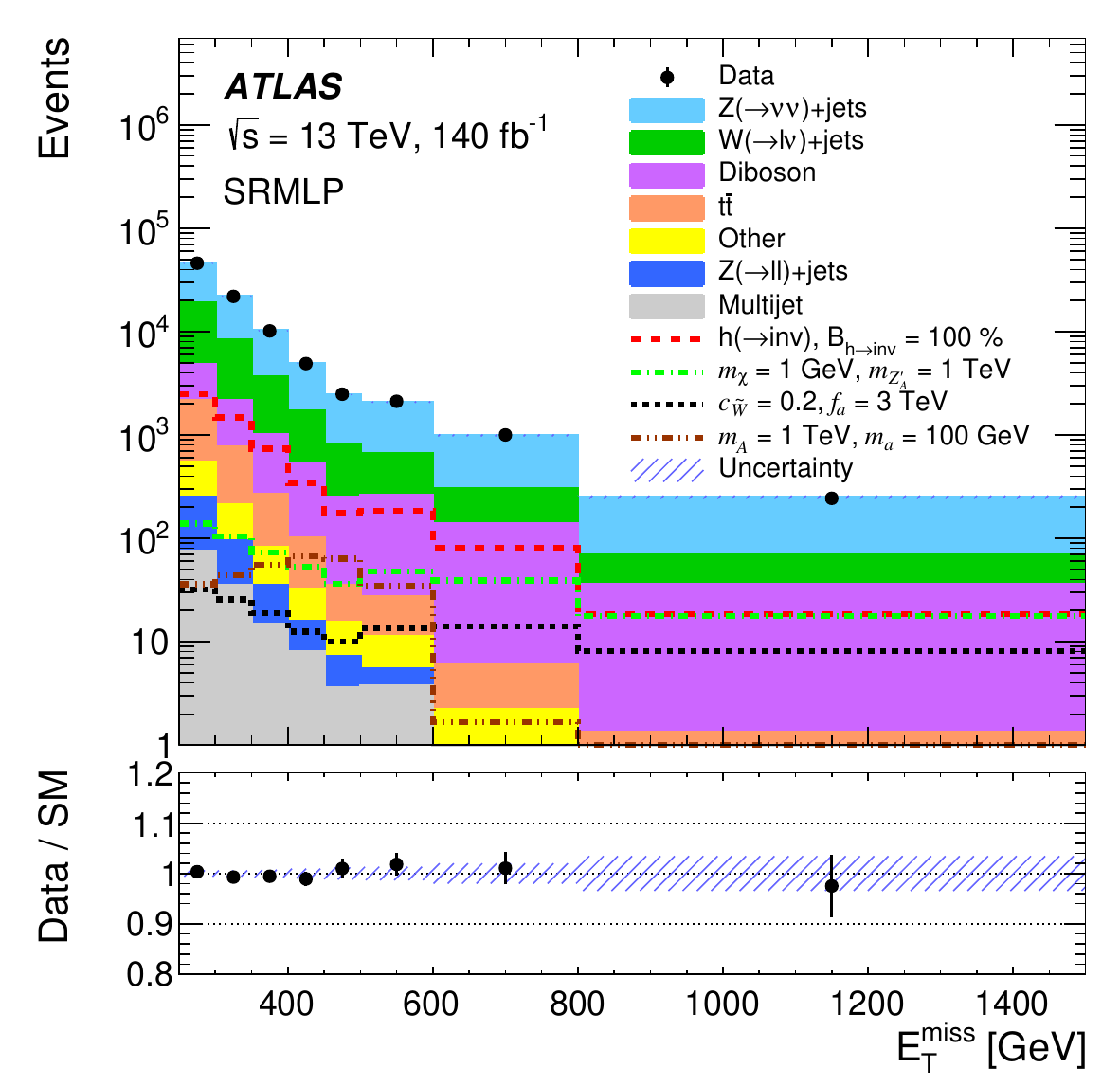}
\caption{}
\end{subfigure}
\begin{subfigure}[t]{0.4\textwidth}
\includegraphics[width=0.99\textwidth]{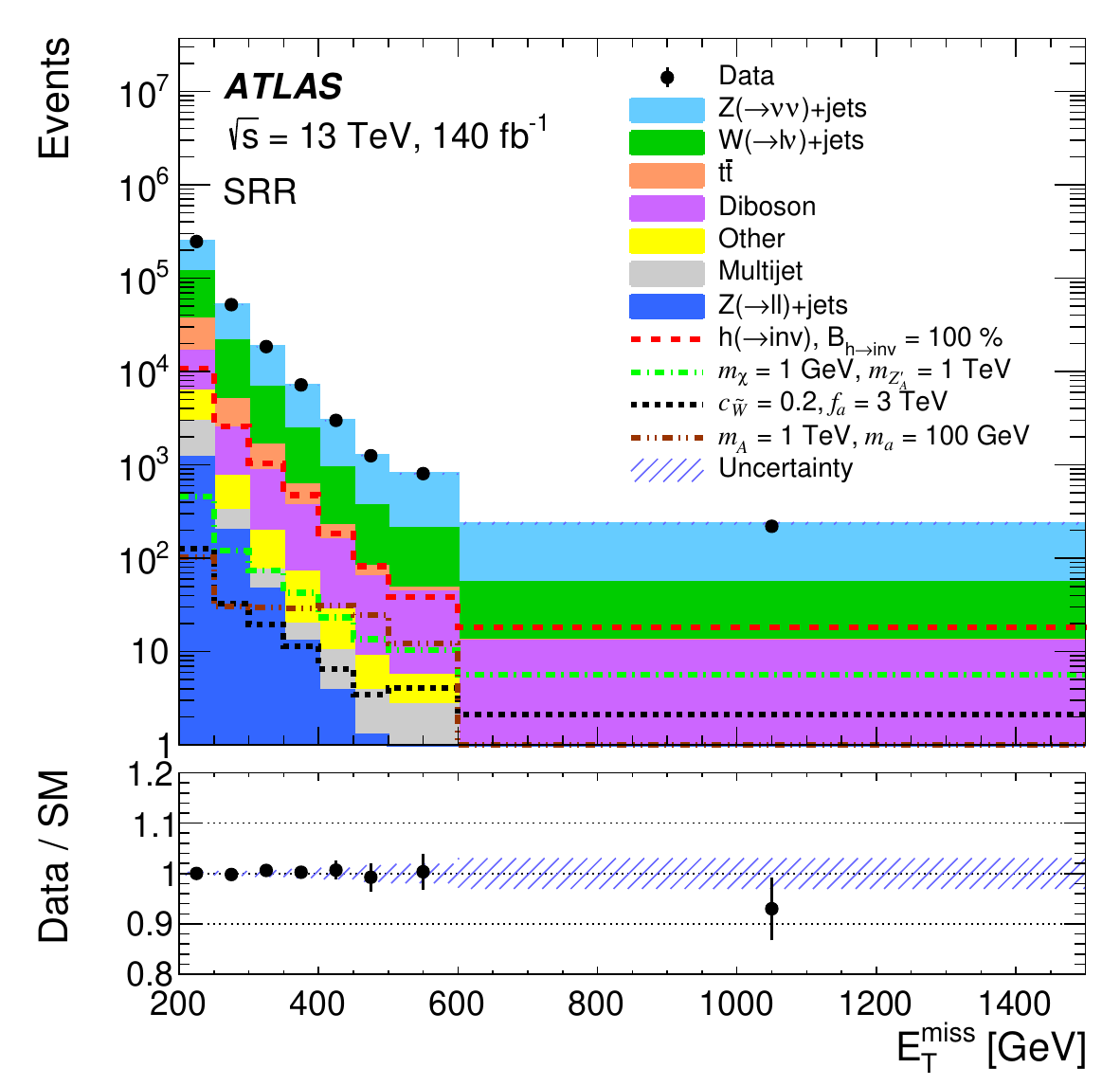}
\caption{}
\end{subfigure}
\caption{Distribution of \met after the background-only fit, in the signal regions: (a) \texttt{SRMHP}, (b) \texttt{SRMLP}, and (c) \texttt{SRR}. The total uncertainty in the background is shown as a shaded band and the error bars on the data points represent the statistical uncertainty. Small background contributions such as $t\bar{t}V$, $Vh$, single-top and triboson processes are grouped into one category and shown as "Other". Example signal distributions with their nominal normalisations are shown, from top to bottom: invisible Higgs, simplified DM model with an axial-vector mediator, ALP and 2HDM+$a$. The last bin contains the overflow.}
\label{fig:bkg_yield}
\end{figure}

Table~\ref{tab:bkg_yield} shows the best-fit background yields in the SRs from the background-only fit, which reflect the consistency of the background model with the observed data. The normalisation factors of the $Z$+jets, $W$+jets, and $t\bar{t}$ backgrounds are listed in Table~\ref{tab:best-fit}. The $t\bar{t}$ merged high purity normalisation factor is lower than that found in the other regions. This is due to a mismodelling of the number of tracks associated with large-$R$ jets in the $t\bar{t}$ simulation~\cite{ATL-PHYS-PUB-2020-017}. This mismodelling leads to a smaller (larger) ratio of data versus MC events in the high (low) purity merged regions. Therefore the $t\bar{t}$ merged high (low) purity normalisation factor is found to be below (above) that of the resolved region, which is not affected by this mismodelling. No significant deviation from the SM background expectation is observed in data. Therefore expected and observed upper limits on signal models are presented in the next section.

\begin{table}[tbp]
\centering
\caption{The expected and observed numbers of events shown separately for each signal region. The background yields and uncertainties are shown after the background-only fit to data. The quoted background uncertainties include both statistical and systematic contributions. The uncertainties in the total background can differ from the squared sum of those on individual components due to correlations of nuisance parameters. Small background contributions such as $t\bar{t}V$, $Vh$, single-top quark and triboson processes are grouped into one category and shown as "Other".}
\begin{tabular}{l
S[table-format = 5.0(3)]
S[table-format = 5.0(4)]
S[table-format = 6.0(4)]
}
\toprule
\midrule
& \multicolumn{3}{c}{Signal Regions} \\
Sample & {\texttt{SRMHP}} & {\texttt{SRMLP}} & {\texttt{SRR}} \\
\midrule
\ensuremath{Z(\rightarrow\nu\nu)+\text{jets}} & 12200 \pm 500 & 54100 \pm 1200 & 180100 \pm 2500 \\
\ensuremath{W(\rightarrow\ell\nu)+\text{jets}} & 6320 \pm 330 & 25600 \pm 700 & 105500 \pm 2000 \\
Diboson & 3700 \pm 700 & 5800 \pm 1500 & 13100 \pm 3000 \\
\ensuremath{t\bar{t}} & 1240 \pm 130 & 2470 \pm 340 & 23600 \pm 1300 \\
Other & 380 \pm 50 & 490 \pm 60 & 3920 \pm 340 \\
\ensuremath{Z(\rightarrow\ell\ell)+\text{jets}} & 55 \pm 6 & 266 \pm 13 & 1480 \pm 40 \\
Multijet & 24 \pm 18 & 140 \pm 100 & 1900 \pm 1400 \\
\midrule
Total background & 23870 \pm 160 & 88880 \pm 300 & 329500 \pm 800 \\
\midrule
Data   & 23861 & 88836 & 329588 \\
\midrule
\bottomrule
\end{tabular}
\label{tab:bkg_yield}
\end{table}

\begin{table}[tbp]
\centering
\caption{The best-fit values of the normalisation factors of the $Z$+jets, $W$+jets, and $t\bar{t}$ backgrounds after the background-only fit.}
\begin{tabular}{c
S[table-format = 1.2(3), table-text-alignment=center]
S[table-format = 1.2(3), table-text-alignment=center]
S[table-format = 1.2(3), table-text-alignment=center]
}
\toprule
\midrule
Normalisation factor & {Merged high purity} & {Merged low purity} & {Resolved} \\
\midrule
\ensuremath{\kappa_{Z\text{\!+jets}}}   & 0.98 \pm 0.09 & 1.03 \pm 0.05 & 0.99 \pm 0.05 \\
\ensuremath{\kappa_{W\text{\!+jets}}}   & 0.95 \pm 0.09 & 0.93 \pm 0.05 & 0.91\pm 0.05 \\
\ensuremath{\kappa_{t\bar{t}}}   & 0.68 \pm 0.12 & 1.03 \pm 0.21 & 0.90 \pm 0.06 \\
\midrule
\bottomrule
\end{tabular}
\label{tab:best-fit}
\end{table}

\clearpage


%
\clearpage
%

\section{Interpretations of the results}
\label{sec:limit}

The results are translated into expected and observed upper limits on the presence of new particles using a simultaneous likelihood fit in both the signal and control regions, and the confidence level ($\mathrm{CL_S}$) modified frequentist approach~\cite{Read:2002hq}.

\subsection{Model-independent exclusion limits}

The model-independent observed and expected upper limits at \SI{95}{\percent}~CL on the number of signal events, $S_{\mathrm{obs}}^{95}$ and $S_\mathrm{{exp}}^{95}$, and on the visible cross-section, $\left<\sigma\right>_{\mathrm{obs}}^{95}$ and $\left<\sigma\right>_\mathrm{{exp}}^{95}$, defined as the product of the production cross-section, acceptance and efficiency, $\sigma \times A \times \varepsilon$, are shown in Figure~\ref{fig:mil}. The limits on the visible cross-section are extracted by dividing the \SI{95}{\percent} CL upper limit on the number of signal events by the integrated luminosity, taking into account the systematic uncertainties in the SM backgrounds and the uncertainty in the integrated luminosity. A fit is performed separately in each SR on the inclusive \MET distribution and is repeated for increasing \MET thresholds indicated by the vertical dashed-grey lines. Values of the visible cross-sections as a function of the \MET threshold that are excluded at \SI{95}{\percent}~CL range from 6.5 to \SI{0.3}{\fb} for \texttt{SRMHP}, from 15.3 to \SI{0.7}{\fb} for \texttt{SRMLH} and from 79.5 to \SI{0.6}{\fb} for \texttt{SRR}.

\begin{figure}[tbp]
\centering
\begin{subfigure}[t]{0.4\textwidth}
\includegraphics[width=0.99\textwidth]{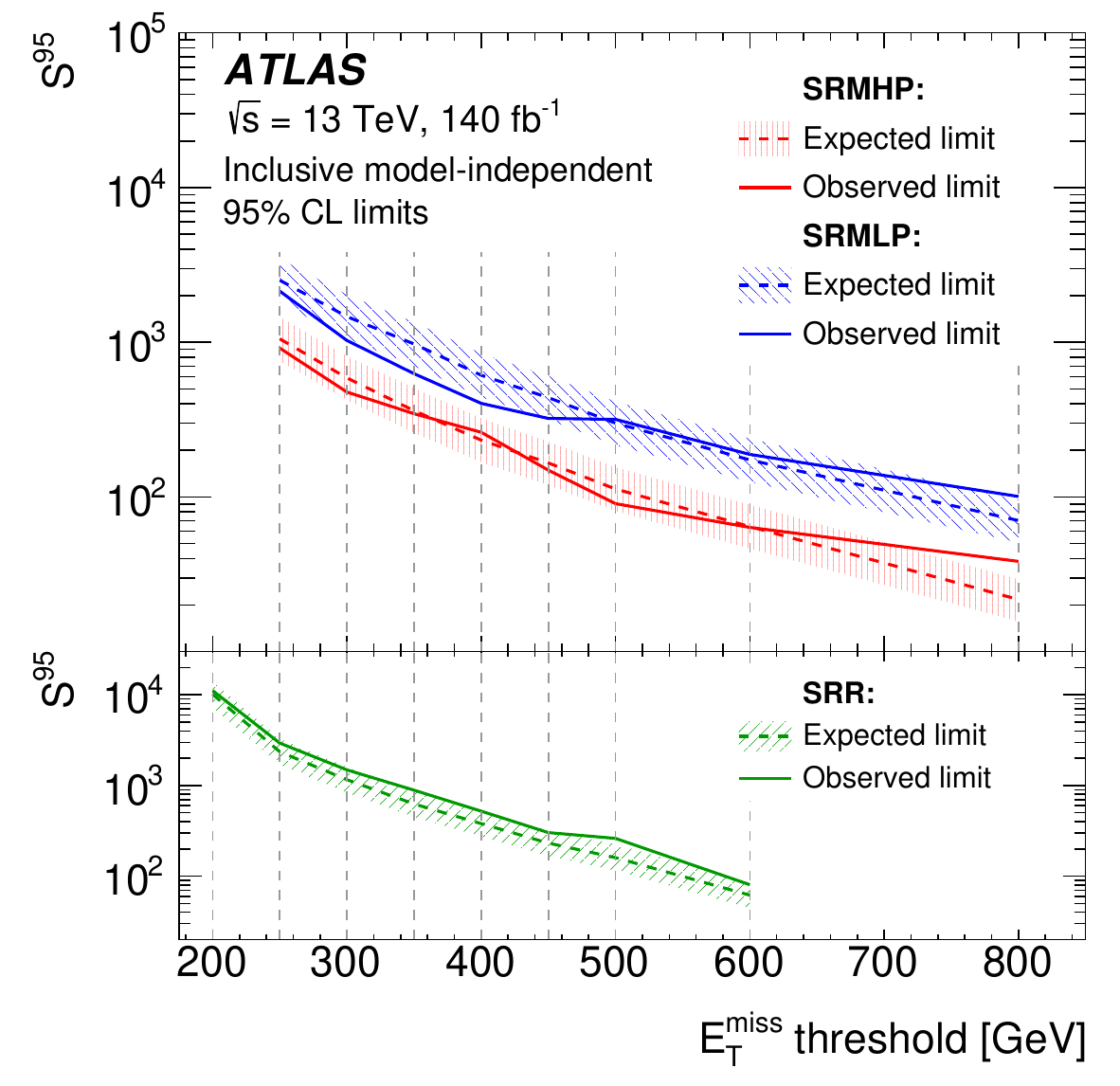}
\caption{}
\end{subfigure}
\begin{subfigure}[t]{0.4\textwidth}
\includegraphics[width=0.99\textwidth]{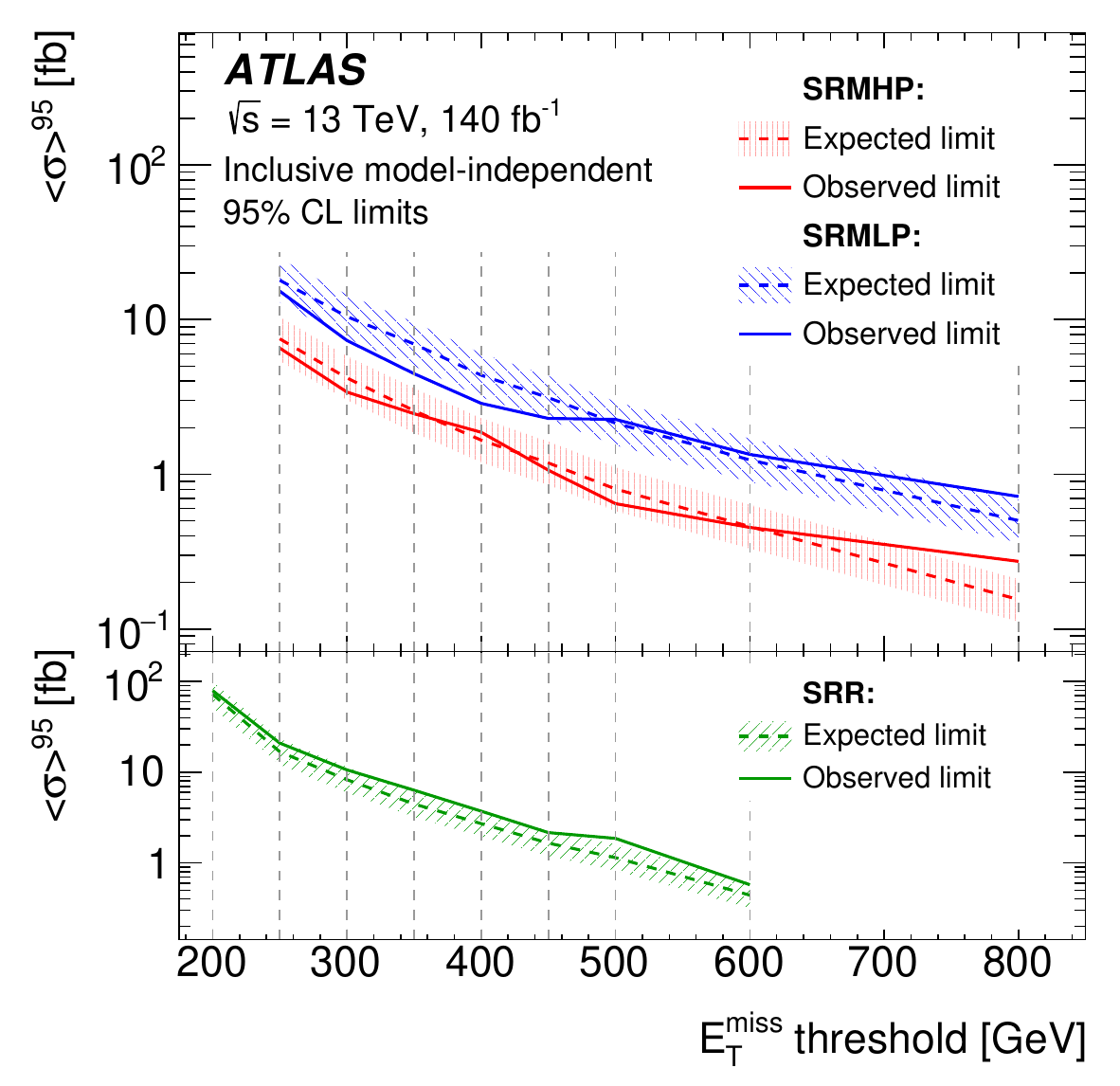}
\caption{}
\end{subfigure}
\caption{Observed and expected \SI{95}{\percent} CL upper limits on (a) the number of signal events, and (b) the visible cross-section for an inclusive model-independent fit. The vertical dashed-grey lines show the starting \MET bin for the range in which the fit is performed.}
\label{fig:mil}
\end{figure}

\subsection{Model-dependent exclusion limits}

Upper limits at 95\% CL are also derived for axion-like particles, the two-Higgs-doublet model with a pseudoscalar, the branching ratio of the Higgs boson to invisible particles, and the simplified DM model with the vector or axial-vector mediator.
\cleardoublepage

\textbf{Axion-like particles}

In the ALP model, the result is expressed in terms of \SI{95}{\percent} CL exclusion limits on the model parameters. Figure~\ref{fig:alps} shows the \SI{95}{\percent} exclusion contour in the $c_{\tilde{W}}$--$f_{a}$ plane for a fixed axion mass of \SI{1}{\MeV}. The exclusion limits do not change significantly for axion masses up to at least \SI{1}{\GeV}. Values of the coupling $c_{\tilde{W}}$ above 0.1 are excluded for $f_{a} = \SI{1}{\TeV}$, and the limit on $c_{\tilde{W}}$ increases linearly with $f_{a}$. In terms of the ratio $c_{\tilde{W}}/f_a$, values above \SI{0.11}{\per\TeV} are excluded. The ALP model is an EFT and becomes invalid for $\hat{s} > f_a^2$ (where $\hat{s} $ corresponds to the invariant mass-squared of the partonic collision). The validity of the effective field theory implementation is verified by applying a suppression factor $f_a^4/\hat{s}^2$~\cite{Brivio:2017ije}. For values of $f_a$ below \SI{2}{TeV}, the signal yields are reduced by up to \SI{50}{\percent} for events with $\hat{s} > f_a^2$ when applying this weighting factor. The reduction is about $5\%$ for $f_a = \SI{2}{TeV}$, while it is negligible for $f_a$ above \SI{3}{TeV}.

\begin{figure}[tbp]
\centering
\includegraphics[width=0.4\textwidth]{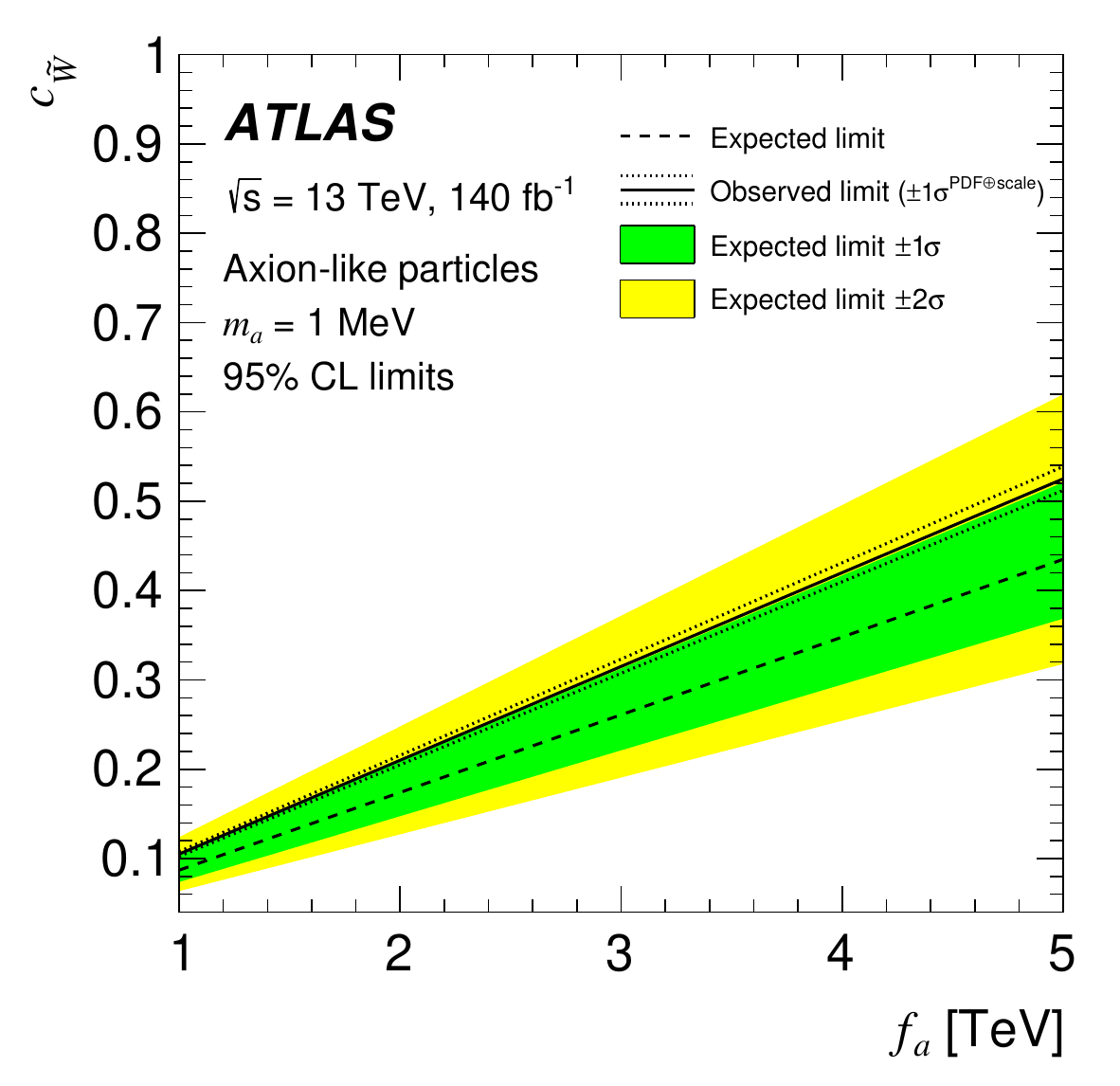}
\caption{Observed (solid line) and expected (dashed line) exclusion upper limits at \SI{95}{\percent} CL on the coupling $c_{\tilde{W}}$ as a function of the effective scale $f_{a}$ for an ALP mass of \SI{1}{\MeV}, with the $\pm1\sigma$ signal theory uncertainties ($\mathrm{PDF}\oplus\mathrm{scale}$) in the observed limit and the $\pm1\sigma$ and $\pm2\sigma$ uncertainties in the expected limit (inner and outer shaded bands). The limits are computed with no suppression of events with $\hat{s} > f_a^2$. The area above the line is excluded at \SI{95}{\percent} CL.}
\label{fig:alps}
\end{figure}

\textbf{Two-Higgs-doublet model with a pseudoscalar}

For the 2HDM+$a$ model, the exclusion limits are derived for all five benchmark scenarios introduced in Section~\ref{sec:data}. The exclusion contours in the $m_{A}$--$m_{a}$ scans with $\tan \beta = 1.0$ for $\sin \theta = 0.35$ and $\sin \theta = 0.7$, which correspond to Scenario 1 introduced in Section~\ref{sec:data}, are shown in Figure~\ref{fig:2hdm1a}. For $\sin \theta = 0.35$~(0.7), the maximum reach is $m_{a}=340\ (420)\ \GeV$ at $m_{A}=\SI{900}{\GeV}$, while values between $m_{A}=520\ (480)\ \GeV$ and $m_{A} = 1100\ (1220)\ \GeV$ are excluded for $m_{a} = \SI{100}{\GeV}$.

\begin{figure}[tbp]
\centering
\begin{subfigure}[t]{0.4\textwidth}
\includegraphics[width=0.99\textwidth]{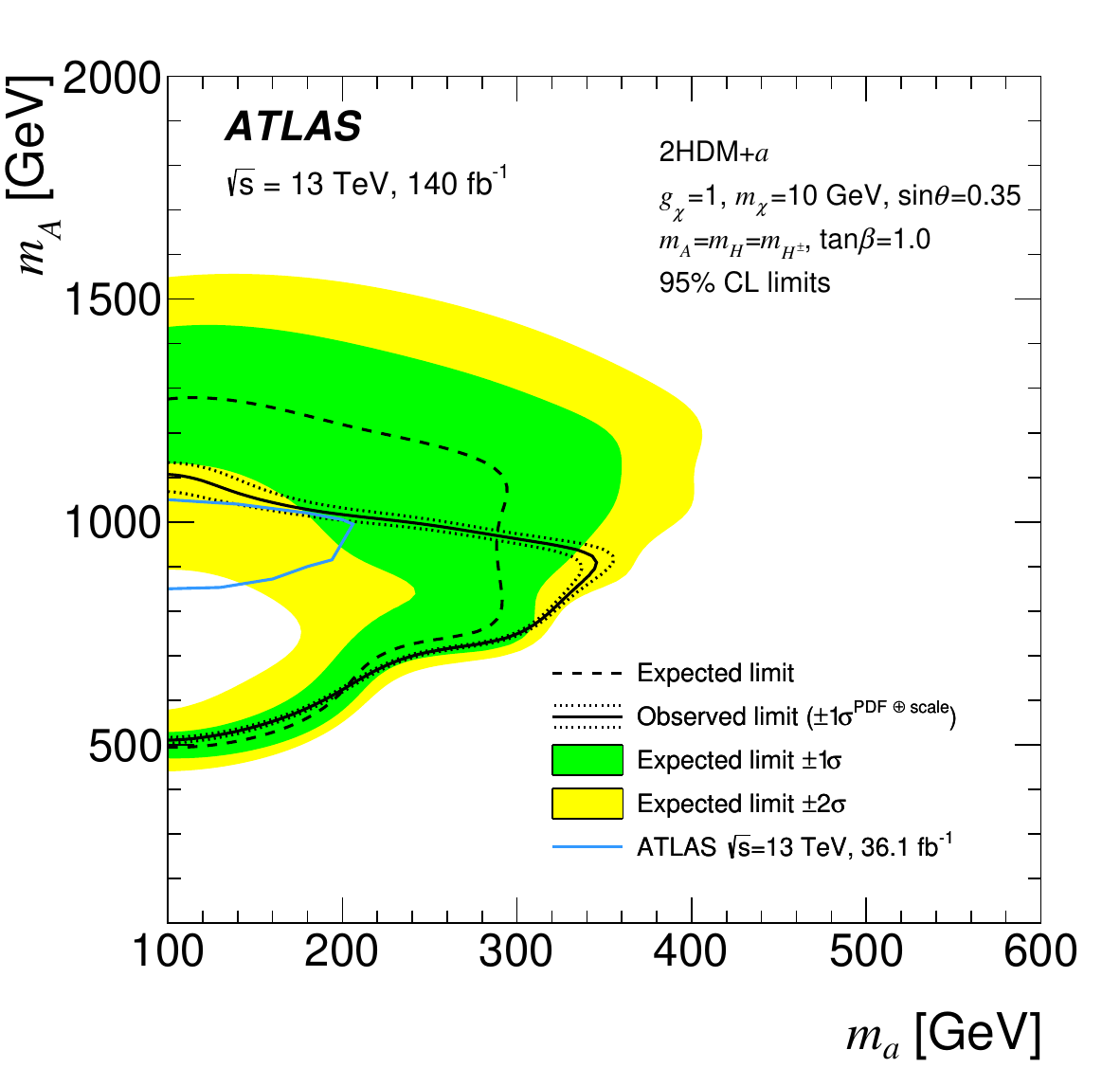}
\caption{}
\end{subfigure}
\begin{subfigure}[t]{0.4\textwidth}
\includegraphics[width=0.99\textwidth]{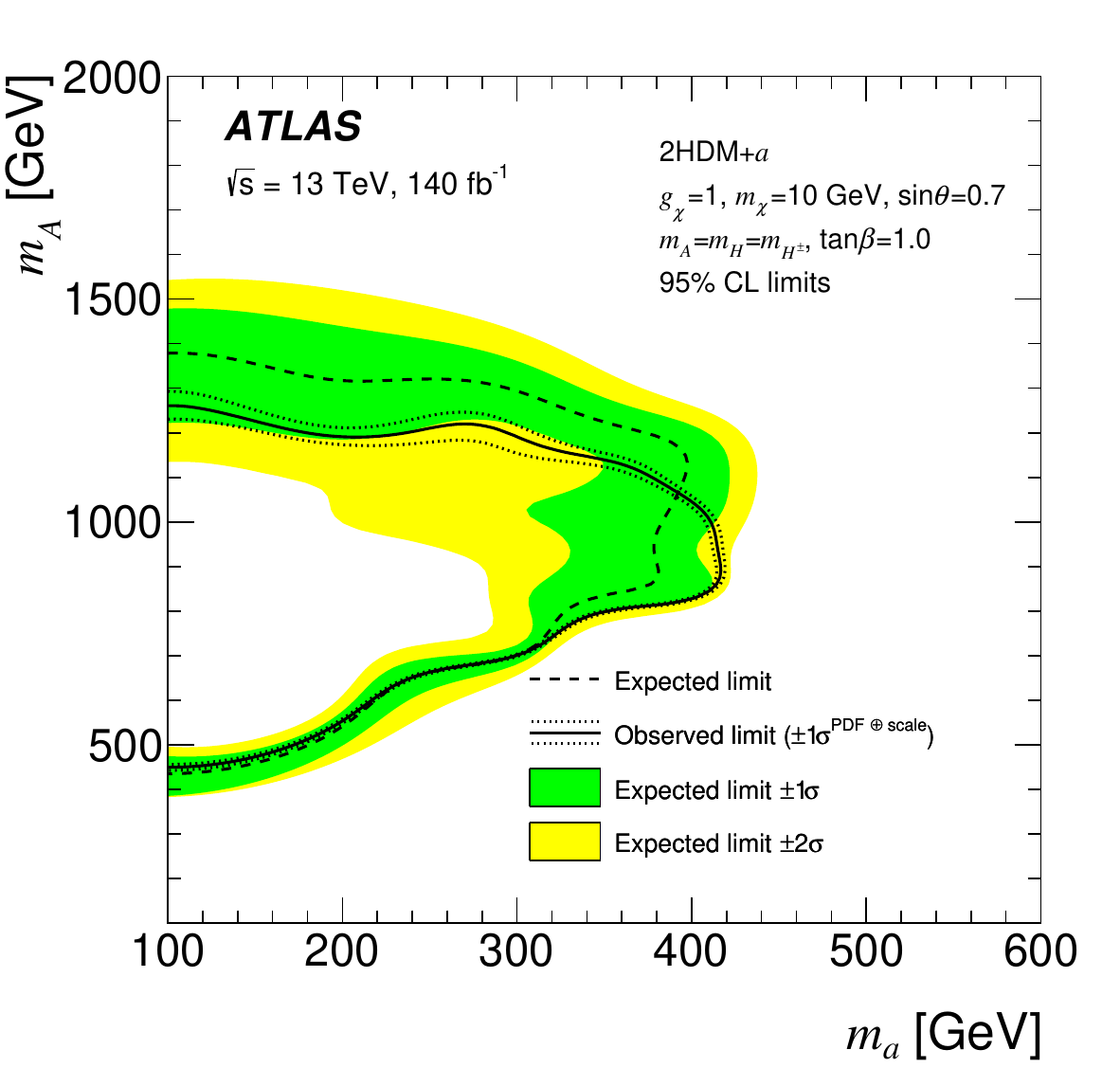}
\caption{}
\end{subfigure}
\caption{Observed (solid line) and expected (dashed line) exclusion upper limits at \SI{95}{\percent} CL in the $m_{A}$--$m_{a}$ plane for the 2HDM+$a$ model, assuming (a) $\sin\theta = 0.35$ and (b) $\sin\theta = 0.7$, with the $\pm1\sigma$ signal theory uncertainties ($\mathrm{PDF}\oplus\mathrm{scale}$) in the observed limit and the $\pm1\sigma$ and $\pm2\sigma$ uncertainties in the expected limit (inner and outer shaded bands). The other model parameters are given in the plot legend. The area inside the closed region is excluded at 95\% CL. The previous ATLAS observed limit obtained with the $\MET+V$ signature for $\sin\theta = 0.35$ is shown with the lighter solid line~\cite{EXOT-2017-32}.}
\label{fig:2hdm1a}
\end{figure}

Figure~\ref{fig:2hdm2b} shows the exclusion limits as a function of $\tan \beta$ and the mass of the pseudo-scalar mediator $m_{a}$ with $m_{A}=\SI{600}{\GeV}$ for both $\theta$ choices: $\sin \theta = 0.35$ and $\sin \theta = 0.7$, which correspond to Scenario 2 introduced in Section~\ref{sec:data}. This search probes values of $m_{a}$ up to $195\ (270)\ \GeV$ for $\tan \beta=0.3$ and $\tan \beta$ up to 12 (2.5) at $m_{a}=\SI{100}{\GeV}$ for $\sin \theta = 0.35$~(0.7).

\begin{figure}[tbp]
\centering
\begin{subfigure}[t]{0.4\textwidth}
\includegraphics[width=0.99\textwidth]{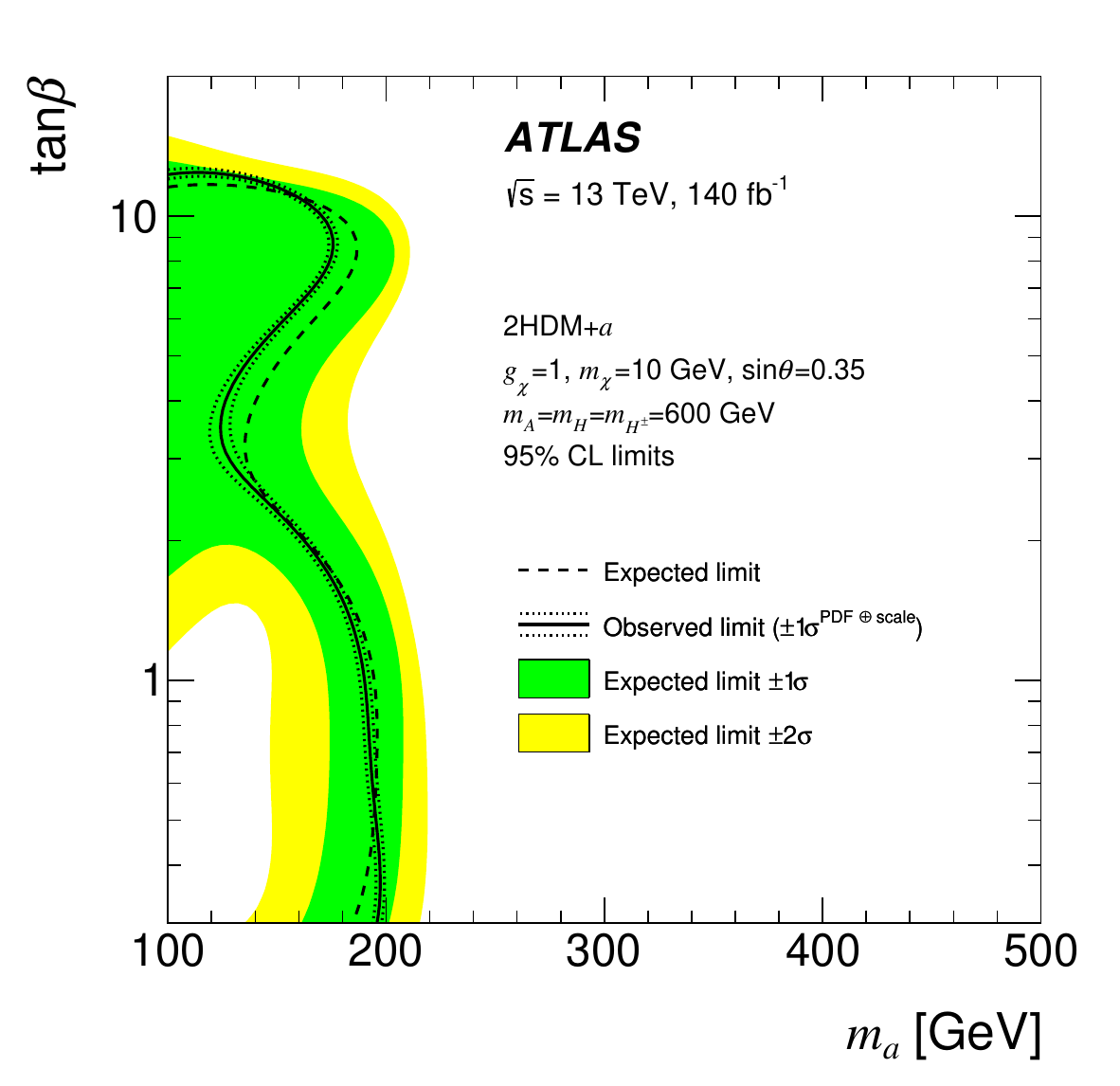}
\caption{}
\end{subfigure}
\begin{subfigure}[t]{0.4\textwidth}
\includegraphics[width=0.99\textwidth]{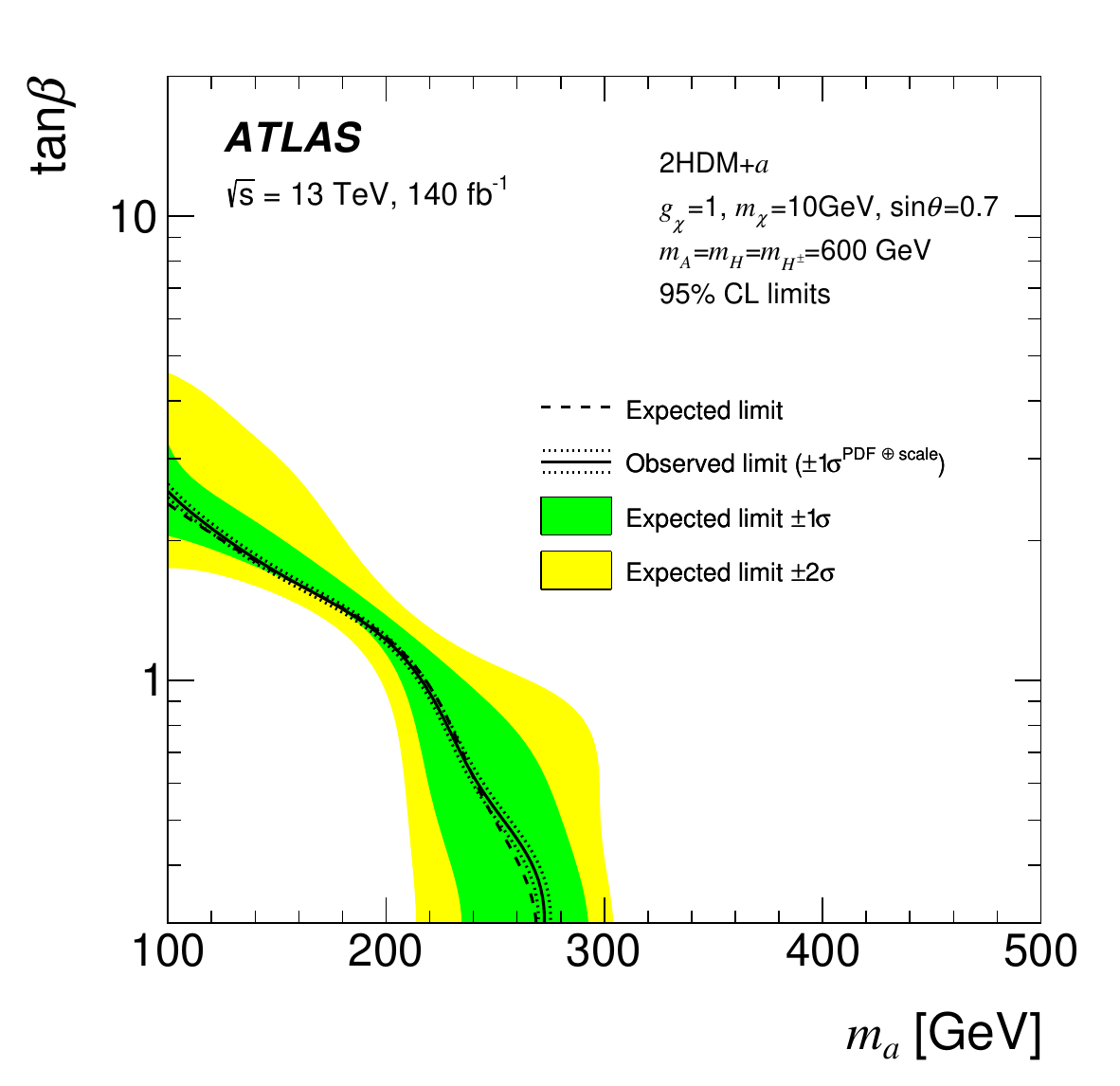}
\caption{}
\end{subfigure}
\caption{Observed (solid line) and expected (dashed line) exclusion upper limits at \SI{95}{\percent} CL in the $\tan \beta$--$m_{a}$ plane for the 2HDM+$a$ model, assuming (a) $\sin \theta = 0.35$ and (b) $\sin \theta = 0.7$, with the $\pm1\sigma$ signal theory uncertainties ($\mathrm{PDF}\oplus\mathrm{scale}$) in the observed limit and the $\pm1\sigma$ and $\pm2\sigma$ uncertainties in the expected limit (inner and outer shaded bands). The other model parameters are given in the plot legend. The area inside the closed region is excluded at 95\% CL.}
\label{fig:2hdm2b}
\end{figure}

An exclusion limit is also derived as a function of $\tan \beta$ and the mass of the pseudo-scalar $m_{A}$ with $m_{a}=\SI{250}{\GeV}$ for both $\theta$ choices: $\sin \theta = 0.35$ and $\sin \theta = 0.7$, which correspond to Scenario 3 introduced in Section~\ref{sec:data}. Figure~\ref{fig:2hdm2c} (a) shows that for $\sin \theta = 0.35$ at $\tan \beta = 20$ masses $m_{A}$ can be excluded between 600 and 1450~GeV, and for $\tan \beta = 0.3$ between 740 and 1450~GeV as well as a small range around 1.7~TeV. The search probes values of $m_{A}$ in the range between 580 and 2010~GeV at $\tan \beta = 0.3$ for $\sin \theta = 0.7$, as shown in Figure~\ref{fig:2hdm2c} (b).

\begin{figure}[tbp]
\centering
\begin{subfigure}[t]{0.4\textwidth}
\includegraphics[width=0.99\textwidth]{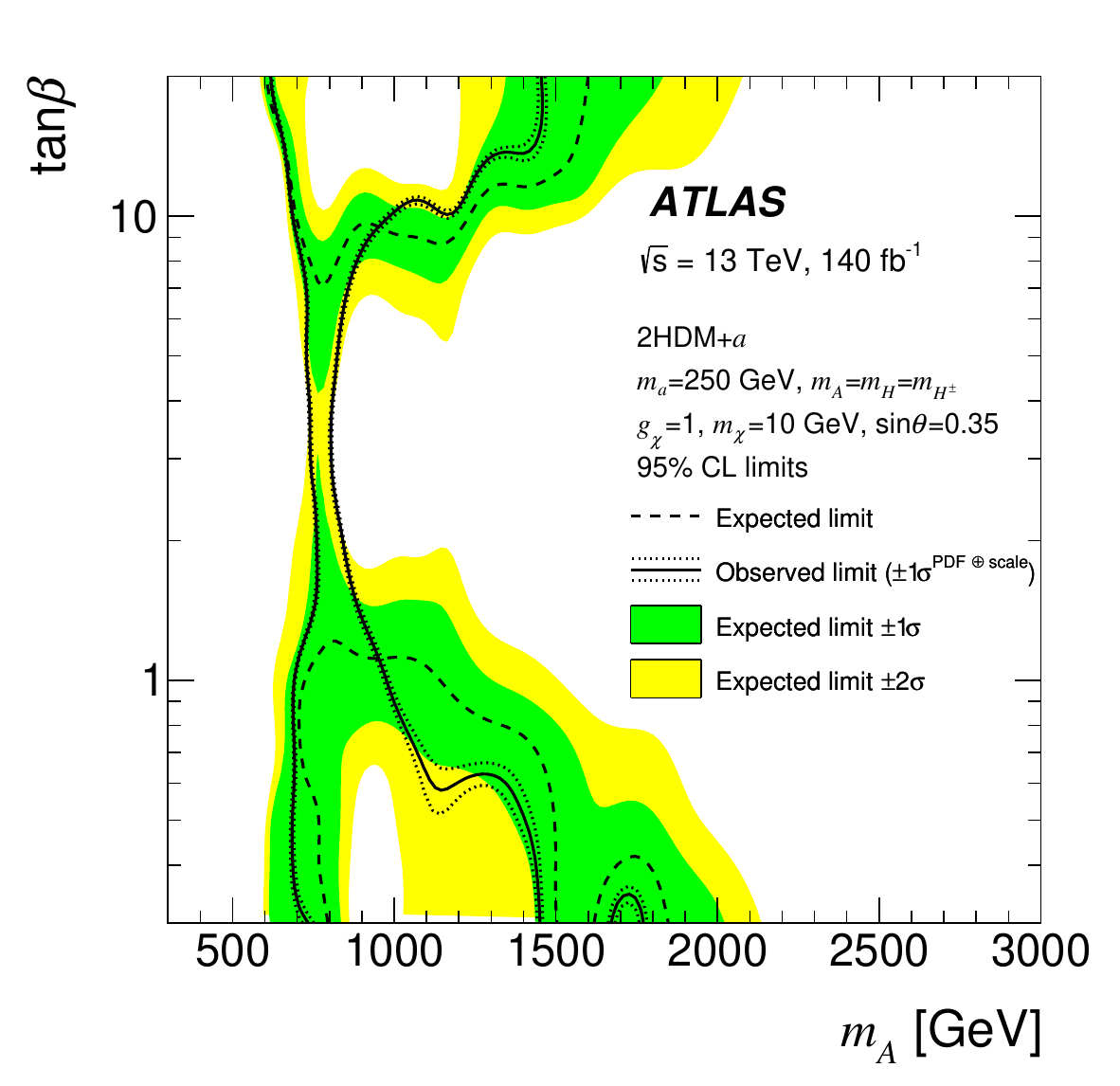}
\caption{}
\end{subfigure}
\begin{subfigure}[t]{0.4\textwidth}
\includegraphics[width=0.99\textwidth]{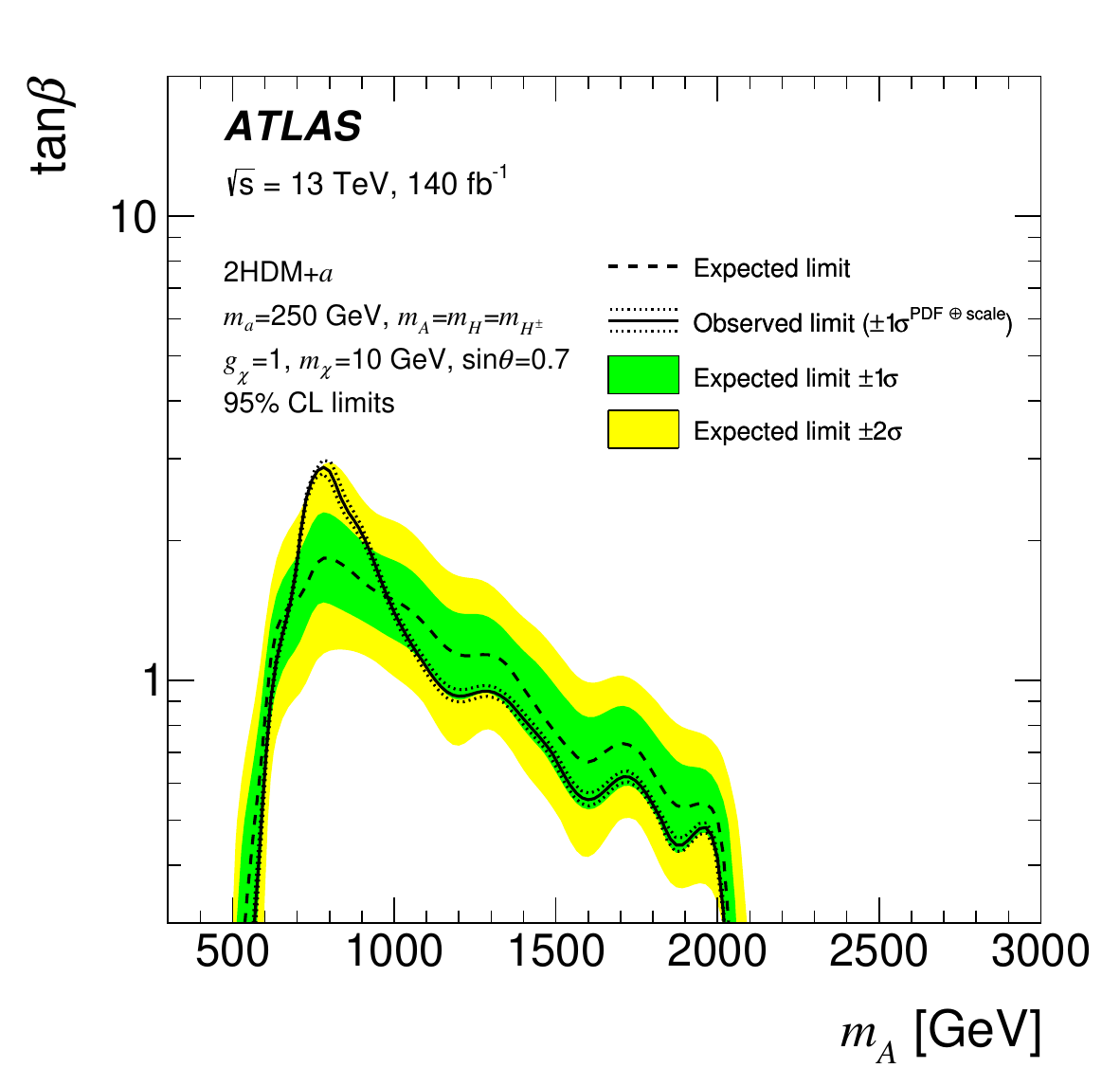}
\caption{}
\end{subfigure}
\caption{Observed (solid line) and expected (dashed line) exclusion upper limits at \SI{95}{\percent} CL in the $\tan \beta$--$m_{A}$ plane for the 2HDM+$a$, assuming (a) $\sin \theta = 0.35$ and (b) $\sin \theta = 0.7$, with the $\pm1\sigma$ signal theory uncertainties ($\mathrm{PDF}\oplus\mathrm{scale}$) in the observed limit and the $\pm1\sigma$ and $\pm2\sigma$ uncertainties in the expected limit (inner and outer shaded bands). The other model parameters are given on the plots. The area inside the closed region is excluded at 95\% CL.}
\label{fig:2hdm2c}
\end{figure}

Figure~\ref{fig:2hdm2d} shows the exclusion limits on the signal strength ($\mu_{sig}$) as a function of $\sin \theta$ for the low mass hypothesis where $m_{a} = \SI{200}{GeV}$ and $m_{A} = \SI{600}{GeV}$, and the high mass hypothesis with $m_{a} = \SI{350}{GeV}$ and $m_{A} = \SI{1000}{GeV}$, which correspond to Scenario 4 introduced in Section~\ref{sec:data}. Values down to 0.6 (1.0) of $\mu_{sig}$ are excluded for the low (high) mass hypothesis at $\sin \theta = 0.6$.

\begin{figure}[tbp]
\centering
\begin{subfigure}[t]{0.4\textwidth}
\includegraphics[width=0.99\textwidth]{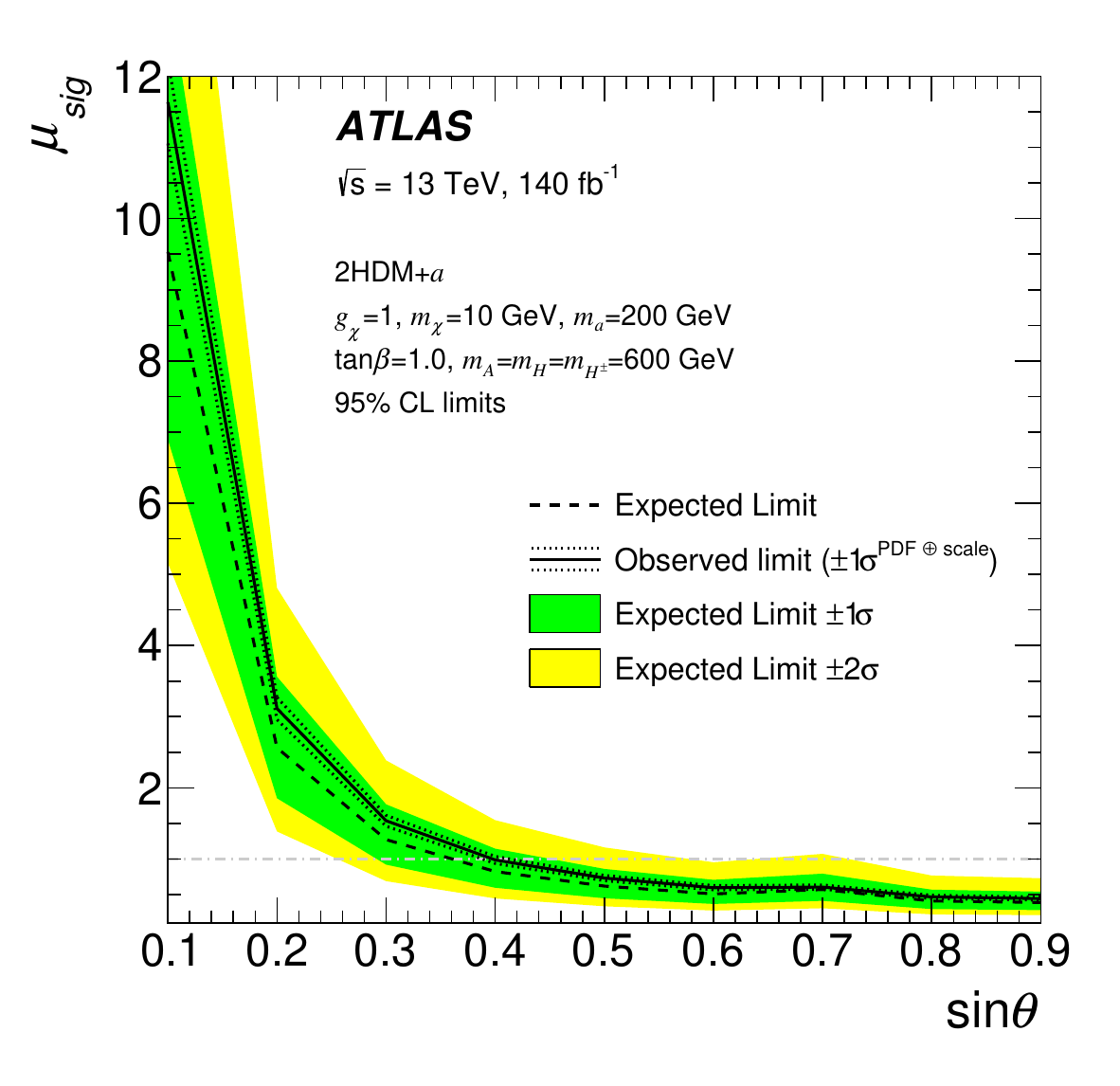}
\caption{}
\end{subfigure}
\begin{subfigure}[t]{0.4\textwidth}
\includegraphics[width=0.99\textwidth]{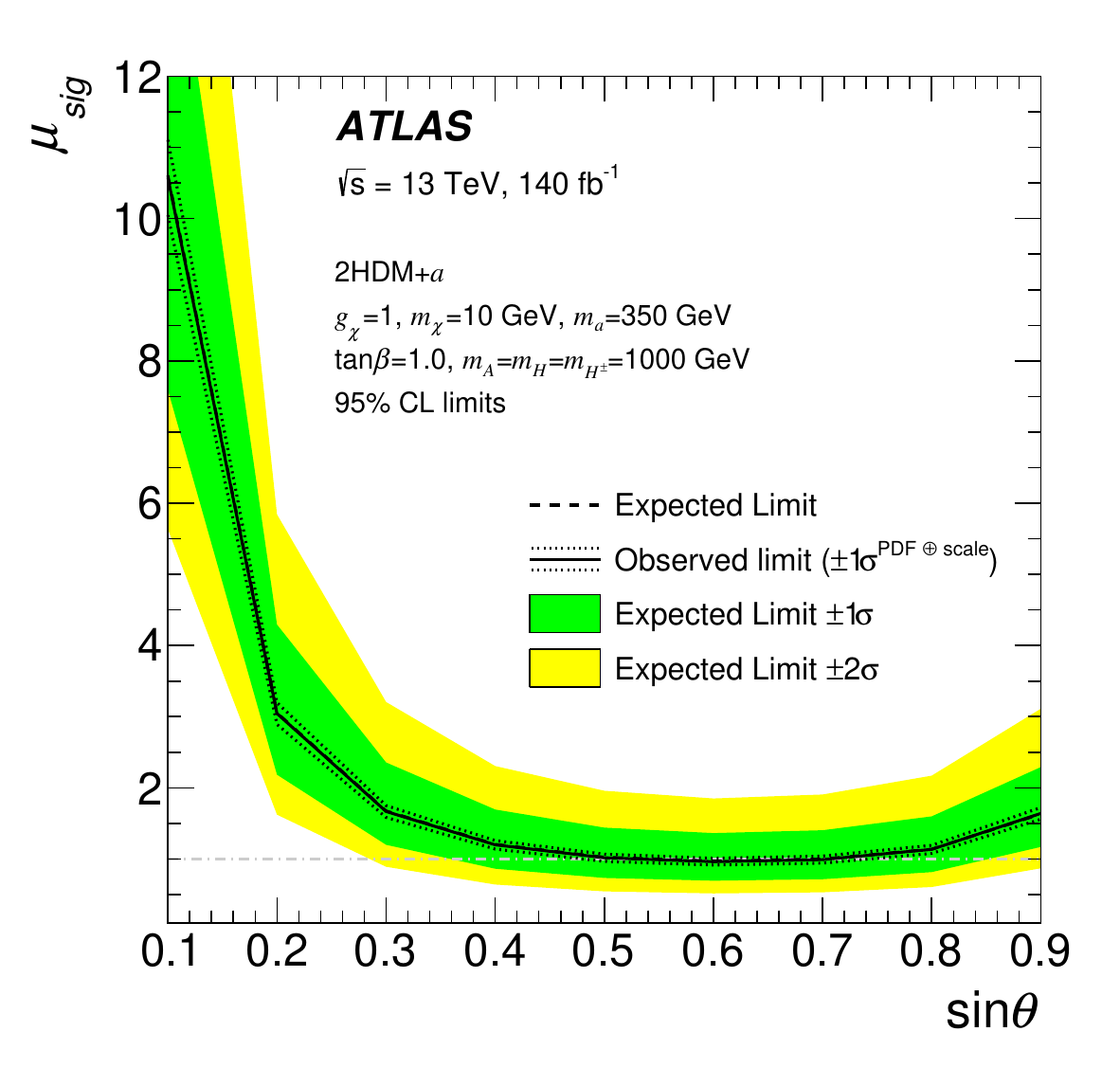}
\caption{}
\end{subfigure}
\caption{Observed (solid line) and expected (dashed line) exclusion upper limits on the signal strength $\mu_{sig}$ at \SI{95}{\percent} CL as a function of $\sin \theta$ for the 2HDM+$a$, with the $\pm1\sigma$ signal theory uncertainties ($\mathrm{PDF}\oplus\mathrm{scale}$) in the observed limit and the $\pm1\sigma$ and $\pm2\sigma$ uncertainties in the expected limit (inner and outer shaded bands): (a) the low mass and (b) high mass hypotheses. The model parameters are given on the plots.}
\label{fig:2hdm2d}
\end{figure}

The exclusion limits on $\mu_{sig}$ as a function of the DM mass $m_{\chi}$ with $\sin \theta = 0.35$ and $\tan \beta = 1.0$ are shown in Figure~\ref{fig:2hdm2e}. The masses of $m_{a}$ and $m_{A}$ are fixed to \SI{400}{\GeV} and \SI{1000}{\GeV}, respectively, corresponding to Scenario 5 introduced in Section~\ref{sec:data}. The result excludes $\mu_{sig}$ down to 2.0 for $m_{\chi} = \SI{100}{\GeV}$.

\begin{figure}[tbp]
\centering
\includegraphics[width=0.4\textwidth]{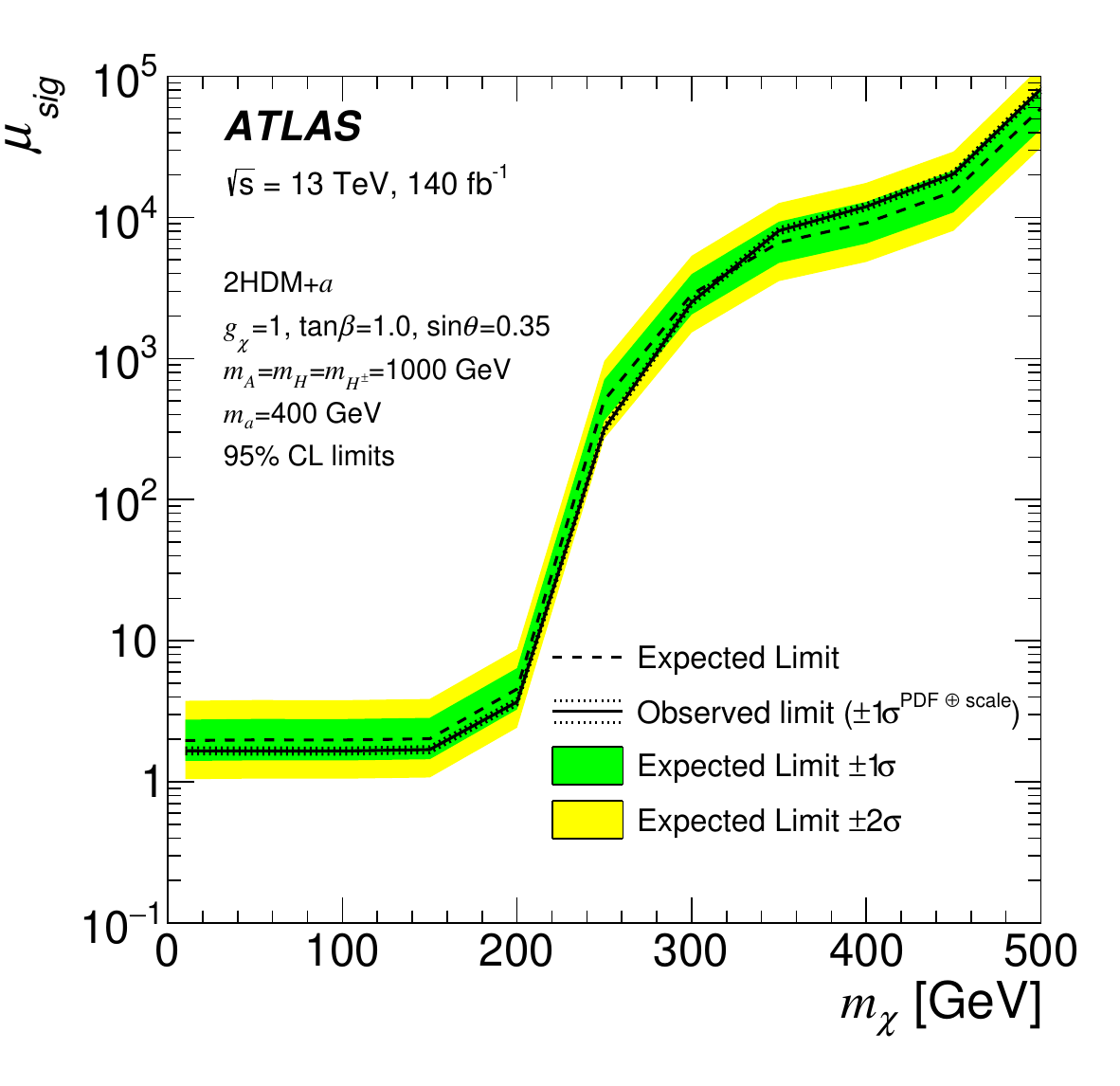}
\caption{Observed (solid line) and expected (dashed line) exclusion upper limits on the signal strength ($\mu_{sig}$) at \SI{95}{\percent} CL as a function of $m_{\chi}$ assuming $m_{a} = \SI{400}{\GeV}$, $m_{A} = \SI{1000}{\GeV}$, $\sin \theta = 0.35$ and $\tan \beta = 1.0$, with the $\pm1\sigma$ signal theory uncertainties ($\mathrm{PDF}\oplus\mathrm{scale}$) in the observed limit and the $\pm1\sigma$ and $\pm2\sigma$ uncertainties in the expected limit (inner and outer shaded bands).}
\label{fig:2hdm2e}
\end{figure}

\textbf{Invisible decaying Higgs boson}

In the search for the invisible decay of the Higgs boson, the result is interpreted as a \SI{95}{\percent} CL upper limit on the branching ratio. The signal yields are dominated by $Vh$ associated production (\SI{69.4}{\percent}) in the \texttt{SRMHP} region. The gluon--gluon fusion production dominates in the \texttt{SRMLP} (\SI{65.6}{\percent}) and \texttt{SRR} (\SI{65.2}{\percent}) regions. As shown in Table~\ref{tab:hinv}, an observed (expected) upper limit of 0.34 ($0.31^{+0.13}_{-0.09}$) is obtained at \SI{95}{\percent} CL on the branching ratio $B_{h\rightarrow inv}$. The observed (expected) upper limits are also shown for the merged and resolved topologies obtained by running separate fits only considering the merged and resolved regions, respectively. The merged topology dominates the sensitivity.

\begin{table}[tbp]
\centering
\caption{Upper limits at \SI{95}{\percent} CL on the branching ratio of the Higgs boson decaying invisibly.}
\begin{tabular}{ccc}
\toprule
\midrule{Limits on $B_{h\rightarrow inv.}$} & {Expected limit} & {Observed limit} \\
\midrule{Merged topology} & $0.34^{+0.14}_{-0.09}$ & 0.38 \\[3pt]
{Resolved topology} & $0.54^{+0.23}_{-0.15}$ & 0.71 \\
\midrule{Combined} & $0.31^{+0.13}_{-0.09}$ & 0.34 \\
\midrule
\bottomrule
\end{tabular}
\label{tab:hinv}
\end{table}

\textbf{Simplified Dark Matter model with a vector or axial-vector mediator}

In the simplified DM model with a vector or axial-vector mediator, the exclusion limits on masses of the DM ($m_{\chi}$) and mediator ($m_{Z'_{V/A}}$) are obtained, assuming Dirac DM with couplings $g_{q}=0.25$, $g_{\chi}=1.0$ and $g_{l}=0$.
Figure~\ref{fig:simplified} shows the observed and expected \SI{95}{\percent} CL exclusion limits for the vector mediator and axial-vector mediator models. The masses of the vector mediator up to \SI{955}{\GeV} are excluded for $m_{\chi} = \SI{1}{\GeV}$. For the axial-vector mediator model, masses up to \SI{965}{\GeV} are excluded for $m_{\chi} = \SI{1}{\GeV}$. Both extend the reach of the previous results obtained with an integrated luminosity of \SI{36.1}{\ifb}~\cite{EXOT-2016-23,EXOT-2017-32}. The masses corresponding to the relic density~\cite{DM_DD} as determined by the Planck~\cite{Planck2018} and WMAP satellites~\cite{2013ApJS..208...19H}, are also presented which cross the excluded region at $m_{Z'_V}$ = \SI{950}{\GeV} and $m_{\chi}$ = \SI{97}{\GeV} for the vector mediator model and $m_{Z'_A}$ = \SI{750}{\GeV} and $m_{\chi}$ = \SI{210}{\GeV} for the axial-vector mediator model. The specific values of the coupling constants are fixed based upon benchmarks provided by the LHC DM Working Group~\cite{DM_DD}. The exclusion limits are found to be strongly dependent on the choice of coupling constants as presented in Ref.~\cite{EXOT-2018-62}.

\begin{figure}[tbp]
\centering
\begin{subfigure}[t]{0.4\textwidth}
\includegraphics[width=0.99\textwidth]{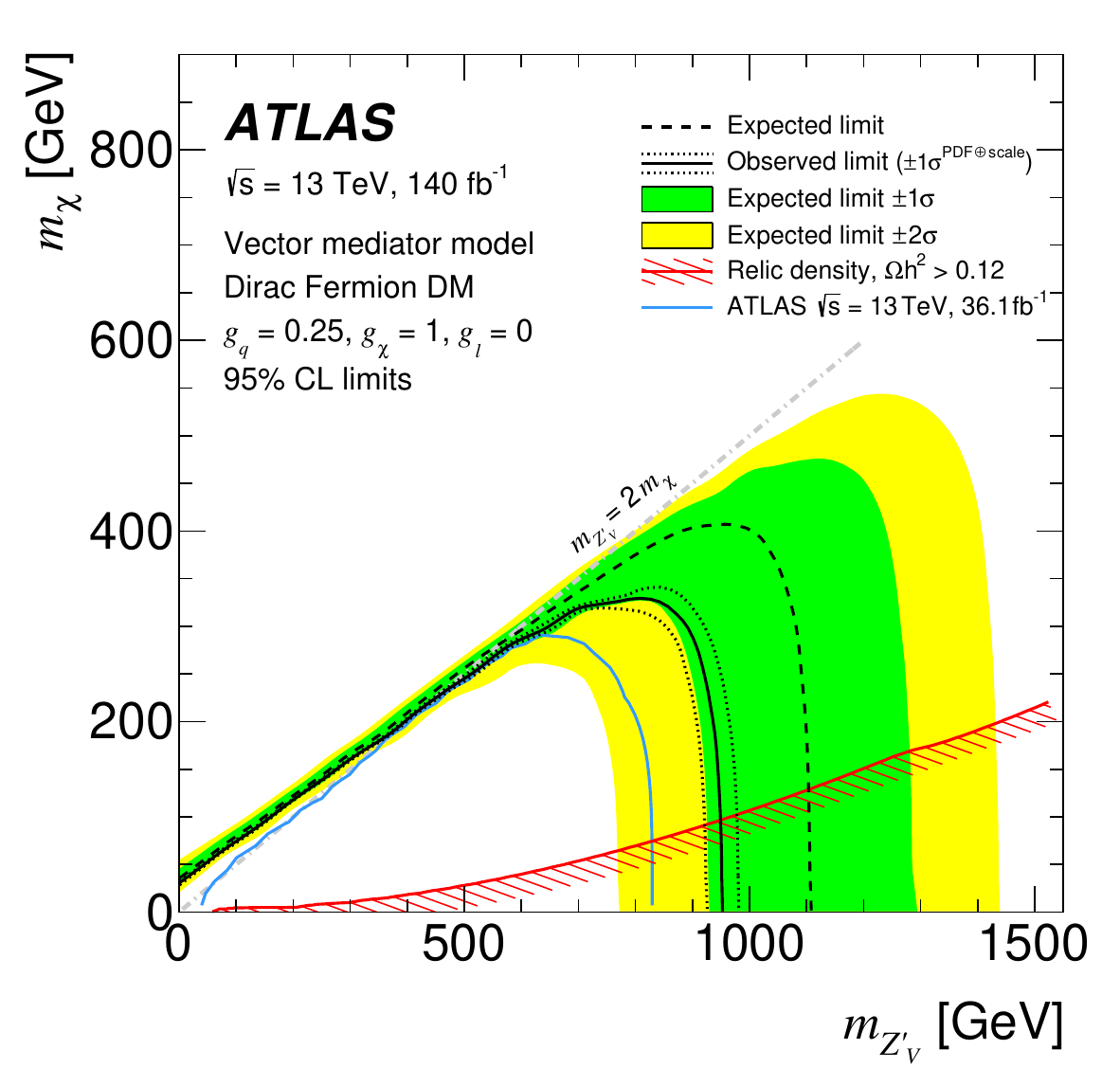}
\caption{}
\end{subfigure}
\begin{subfigure}[t]{0.4\textwidth}
\includegraphics[width=0.99\textwidth]{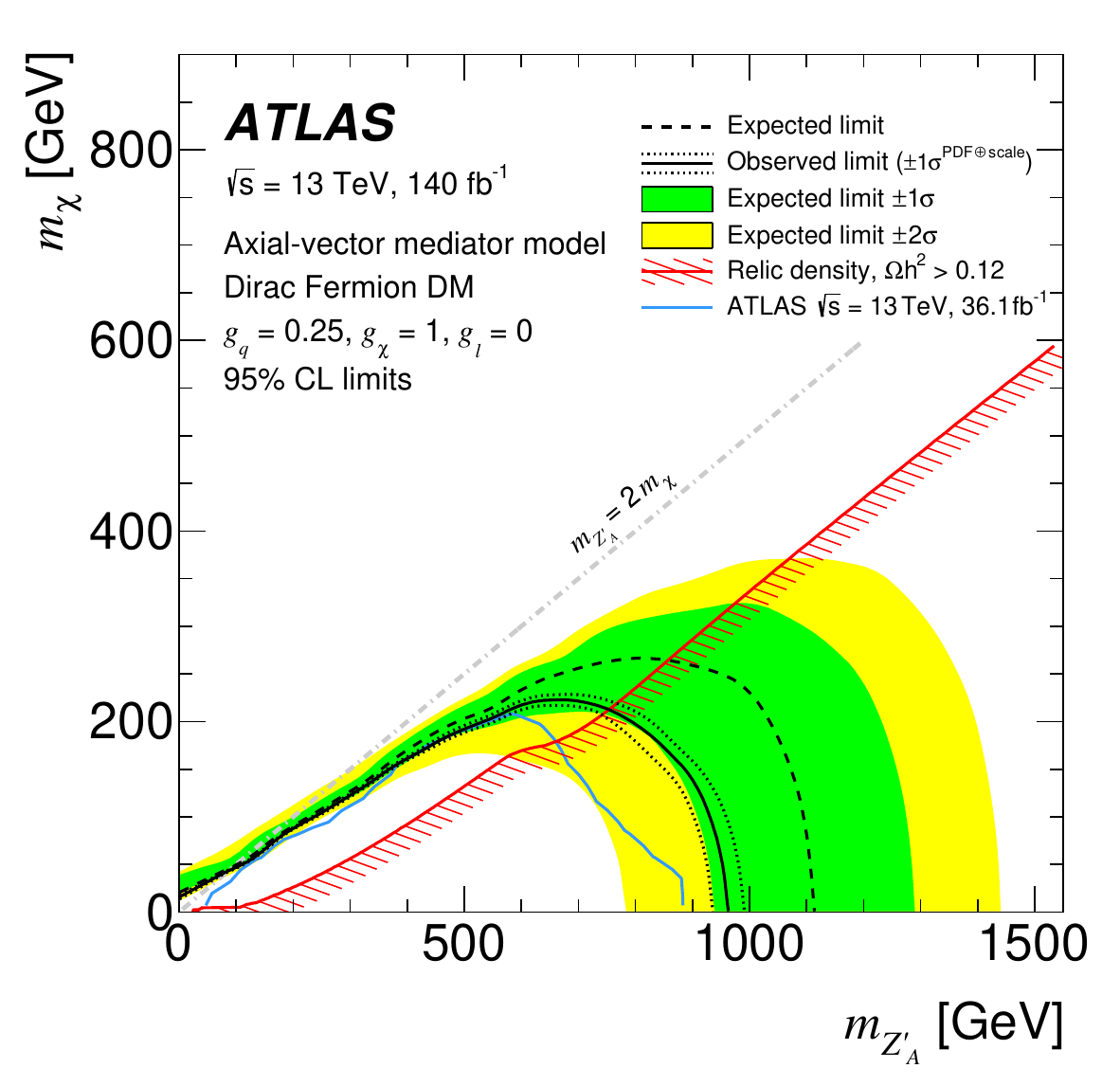}
\caption{}
\end{subfigure}
\caption{Exclusion contours at \SI{95}{\percent} CL on the DM and mediator masses in the simplified DM model with (a) a vector or (b) an axial-vector mediator. The solid (dashed) curve shows the observed (expected) limit, with the $\pm1\sigma$ signal theory uncertainties ($\mathrm{PDF}\oplus\mathrm{scale}$) in the observed limit and the $\pm1\sigma$ and $\pm2\sigma$ uncertainties in the expected limit (inner and outer shaded bands). The previous ATLAS observed limit is shown with the lighter solid curve~\cite{EXOT-2016-23,EXOT-2017-32}. The hashed curves show the set of points for which the expected relic density is consistent with the Planck and WMAP measurements (i.e.\ $\Omega h^2 = 0.12$), as computed with {\normalfont \scshape MadDM}~\cite{Backovic:2015cra}. The area on the hashed side of this curve corresponds to predicted values of the relic density abundance larger than the measured one ($\Omega h^2 > 0.12$).}
\label{fig:simplified}
\end{figure}


%

\section{Conclusion}
\label{sec:summary}

A search for new particles has been performed in events with missing transverse momentum and either a large-$R$ jet or a pair of small-$R$ jets compatible with the hadronic decay of a $W$ or $Z$ boson using 140~\ifb of proton--proton collision data at $\sqrt{s}=\SI{13}{\TeV}$ recorded with the ATLAS detector at the LHC between 2015 and 2018. In addition to the increased data sample size and the use of improved criteria and reduced uncertainties for reconstructed objects, the analysis implements several improvements in the selection of events, for instance improved $b$- and $V$-tagging and the $\tau$-lepton veto, to provide better sensitivity. No significant excess is observed above the expected SM background. %

Model-independent upper exclusion limits are derived in the range from \SI{79.5}{\fb} to \SI{0.3}{\fb} at \SI{95}{\percent} CL for increasing missing transverse momentum thresholds for all the search regions. Exclusion limits are derived for the coupling $c_{\tilde{W}}$ in the ALP model; couplings $c_{\tilde{W}}$ above 0.1 are excluded for $f_{a} = \SI{1}{\TeV}$. Exclusion limits are also derived for the two-Higgs-doublet model with a pseudo-scalar mediator for all five benchmark scenarios. Masses of the pseudo-scalar mediator $m_{a}$ are excluded up to 340 (420)~GeV for $m_{A} = \SI{900}{\GeV}$, sin $\theta =0.35$ (0.7), and tan $\beta =1.0$.
The observed upper limit on the branching ratio of the Higgs boson to invisible final states, $B_{h\rightarrow inv}$, is found to be 0.34 at \SI{95}{\percent} CL; the corresponding expected limit is 0.31. The expected limit is a factor of 1.9 better than the previous ATLAS $\MET+V$ search based on 36 \ifb. In the simplified model in which the DM is produced via an $s$-channel exchange of a vector mediator, mediator masses up to \SI{955}{\GeV} are excluded for a DM mass of \SI{1}{\GeV} and a fixed choice of couplings; in the case of an axial-vector mediator, the exclusion region covers mediator masses up to \SI{965}{\GeV}.


%
\section*{Acknowledgements}
%

%
%

%
%

We thank CERN for the very successful operation of the LHC and its injectors, as well as the support staff at
CERN and at our institutions worldwide without whom ATLAS could not be operated efficiently.

The crucial computing support from all WLCG partners is acknowledged gratefully, in particular from CERN, the ATLAS Tier-1 facilities at TRIUMF/SFU (Canada), NDGF (Denmark, Norway, Sweden), CC-IN2P3 (France), KIT/GridKA (Germany), INFN-CNAF (Italy), NL-T1 (Netherlands), PIC (Spain), RAL (UK) and BNL (USA), the Tier-2 facilities worldwide and large non-WLCG resource providers. Major contributors of computing resources are listed in Ref.~\cite{ATL-SOFT-PUB-2023-001}.

We gratefully acknowledge the support of ANPCyT, Argentina; YerPhI, Armenia; ARC, Australia; BMWFW and FWF, Austria; ANAS, Azerbaijan; CNPq and FAPESP, Brazil; NSERC, NRC and CFI, Canada; CERN; ANID, Chile; CAS, MOST and NSFC, China; Minciencias, Colombia; MEYS CR, Czech Republic; DNRF and DNSRC, Denmark; IN2P3-CNRS and CEA-DRF/IRFU, France; SRNSFG, Georgia; BMBF, HGF and MPG, Germany; GSRI, Greece; RGC and Hong Kong SAR, China; ISF and Benoziyo Center, Israel; INFN, Italy; MEXT and JSPS, Japan; CNRST, Morocco; NWO, Netherlands; RCN, Norway; MNiSW, Poland; FCT, Portugal; MNE/IFA, Romania; MESTD, Serbia; MSSR, Slovakia; ARIS and MVZI, Slovenia; DSI/NRF, South Africa; MICIU/AEI, Spain; SRC and Wallenberg Foundation, Sweden; SERI, SNSF and Cantons of Bern and Geneva, Switzerland; NSTC, Taipei; TENMAK, T\"urkiye; STFC/UKRI, United Kingdom; DOE and NSF, United States of America.

Individual groups and members have received support from BCKDF, CANARIE, CRC and DRAC, Canada; CERN-CZ, FORTE and PRIMUS, Czech Republic; COST, ERC, ERDF, Horizon 2020, ICSC-NextGenerationEU and Marie Sk{\l}odowska-Curie Actions, European Union; Investissements d'Avenir Labex, Investissements d'Avenir Idex and ANR, France; DFG and AvH Foundation, Germany; Herakleitos, Thales and Aristeia programmes co-financed by EU-ESF and the Greek NSRF, Greece; BSF-NSF and MINERVA, Israel; Norwegian Financial Mechanism 2014-2021, Norway; NCN and NAWA, Poland; La Caixa Banking Foundation, CERCA Programme Generalitat de Catalunya and PROMETEO and GenT Programmes Generalitat Valenciana, Spain; G\"{o}ran Gustafssons Stiftelse, Sweden; The Royal Society and Leverhulme Trust, United Kingdom.

In addition, individual members wish to acknowledge support from CERN: European Organization for Nuclear Research (CERN PJAS); Chile: Agencia Nacional de Investigaci\'on y Desarrollo (FONDECYT 1190886, FONDECYT 1230812, FONDECYT 1230987); China: Chinese Ministry of Science and Technology (MOST-2023YFA1605700), National Natural Science Foundation of China (NSFC - 12175119, NSFC 12275265, NSFC-12075060); Czech Republic: Czech Science Foundation (GACR - 24-11373S), Ministry of Education Youth and Sports (FORTE CZ.02.01.01/00/22\_008/0004632), PRIMUS Research Programme (PRIMUS/21/SCI/017); EU: H2020 European Research Council (ERC - 101002463); European Union: European Research Council (ERC - 948254, ERC 101089007), Horizon 2020 Framework Programme (MUCCA - CHIST-ERA-19-XAI-00), European Union, Future Artificial Intelligence Research (FAIR-NextGenerationEU PE00000013), Italian Center for High Performance Computing, Big Data and Quantum Computing (ICSC, NextGenerationEU); France: Agence Nationale de la Recherche (ANR-20-CE31-0013, ANR-21-CE31-0013, ANR-21-CE31-0022, ANR-22-EDIR-0002), Investissements d'Avenir Labex (ANR-11-LABX-0012); Germany: Baden-Württemberg Stiftung (BW Stiftung-Postdoc Eliteprogramme), Deutsche Forschungsgemeinschaft (DFG - 469666862, DFG - CR 312/5-2); Italy: Istituto Nazionale di Fisica Nucleare (ICSC, NextGenerationEU), Ministero dell'Università e della Ricerca (PRIN - 20223N7F8K - PNRR M4.C2.1.1); Japan: Japan Society for the Promotion of Science (JSPS KAKENHI JP21H05085, JSPS KAKENHI JP22H01227, JSPS KAKENHI JP22H04944, JSPS KAKENHI JP22KK0227); Netherlands: Netherlands Organisation for Scientific Research (NWO Veni 2020 - VI.Veni.202.179); Norway: Research Council of Norway (RCN-314472); Poland: Ministry of Science and Higher Education (IDUB AGH, POB8, D4 no 9722), Polish National Agency for Academic Exchange (PPN/PPO/2020/1/00002/U/00001), Polish National Science Centre (NCN 2021/42/E/ST2/00350, NCN OPUS nr 2022/47/B/ST2/03059, NCN UMO-2019/34/E/ST2/00393, UMO-2020/37/B/ST2/01043, UMO-2021/40/C/ST2/00187, UMO-2022/47/O/ST2/00148, UMO-2023/49/B/ST2/04085); Slovenia: Slovenian Research Agency (ARIS grant J1-3010); Spain: Generalitat Valenciana (Artemisa, FEDER, IDIFEDER/2018/048), Ministry of Science and Innovation (MCIN \& NextGenEU PCI2022-135018-2, MICIN \& FEDER PID2021-125273NB, RYC2019-028510-I, RYC2020-030254-I, RYC2021-031273-I, RYC2022-038164-I), PROMETEO and GenT Programmes Generalitat Valenciana (CIDEGENT/2019/023, CIDEGENT/2019/027); Sweden: Swedish Research Council (Swedish Research Council 2023-04654, VR 2018-00482, VR 2022-03845, VR 2022-04683, VR 2023-03403, VR grant 2021-03651), Knut and Alice Wallenberg Foundation (KAW 2018.0157, KAW 2018.0458, KAW 2019.0447, KAW 2022.0358); Switzerland: Swiss National Science Foundation (SNSF - PCEFP2\_194658); United Kingdom: Leverhulme Trust (Leverhulme Trust RPG-2020-004), Royal Society (NIF-R1-231091); United States of America: U.S. Department of Energy (ECA DE-AC02-76SF00515), Neubauer Family Foundation.

%
%


%
%
%
%
%
%
%

%

%
%
%
%

%
%

%
%
%
\printbibliography
\clearpage
\input{atlas_authlist}
%

%
%

%

%
%
%
%

\end{document}

%% file: atlas_authlist.tex
 
\begin{flushleft}
\hypersetup{urlcolor=black}
{\Large The ATLAS Collaboration}

\bigskip

\AtlasOrcid[0000-0002-6665-4934]{G.~Aad}$^\textrm{\scriptsize 104}$,
\AtlasOrcid[0000-0001-7616-1554]{E.~Aakvaag}$^\textrm{\scriptsize 17}$,
\AtlasOrcid[0000-0002-5888-2734]{B.~Abbott}$^\textrm{\scriptsize 123}$,
\AtlasOrcid[0000-0002-0287-5869]{S.~Abdelhameed}$^\textrm{\scriptsize 119a}$,
\AtlasOrcid[0000-0002-1002-1652]{K.~Abeling}$^\textrm{\scriptsize 56}$,
\AtlasOrcid[0000-0001-5763-2760]{N.J.~Abicht}$^\textrm{\scriptsize 50}$,
\AtlasOrcid[0000-0002-8496-9294]{S.H.~Abidi}$^\textrm{\scriptsize 30}$,
\AtlasOrcid[0009-0003-6578-220X]{M.~Aboelela}$^\textrm{\scriptsize 45}$,
\AtlasOrcid[0000-0002-9987-2292]{A.~Aboulhorma}$^\textrm{\scriptsize 36e}$,
\AtlasOrcid[0000-0001-5329-6640]{H.~Abramowicz}$^\textrm{\scriptsize 155}$
% ...
\bigskip
\\

% ...
$^{17}$Department for Physics and Technology, University of Bergen, Bergen; Norway.\\
% ...
$^{30}$Physics Department, Brookhaven National Laboratory, Upton NY; United States of America.\\
% ...
$^{36}$$^{(a)}$Facult\'e des Sciences Ain Chock, Universit\'e Hassan II de Casablanca;$^{(b)}$Facult\'{e} des Sciences, Universit\'{e} Ibn-Tofail, K\'{e}nitra;$^{(c)}$Facult\'e des Sciences Semlalia, Universit\'e Cadi Ayyad, LPHEA-Marrakech;$^{(d)}$LPMR, Facult\'e des Sciences, Universit\'e Mohamed Premier, Oujda;$^{(e)}$Facult\'e des sciences, Universit\'e Mohammed V, Rabat;$^{(f)}$Institute of Applied Physics, Mohammed VI Polytechnic University, Ben Guerir; Morocco.\\
% ...
$^{45}$Physics Department, Southern Methodist University, Dallas TX; United States of America.\\
% ...
$^{50}$Fakult\"{a}t Physik , Technische Universit{\"a}t Dortmund, Dortmund; Germany.\\
% ...
$^{56}$II. Physikalisches Institut, Georg-August-Universit\"{a}t G\"ottingen, G\"ottingen; Germany.\\
% ...
$^{104}$CPPM, Aix-Marseille Universit\'e, CNRS/IN2P3, Marseille; France.\\
% ...
$^{119}$$^{(a)}$New York University Abu Dhabi, Abu Dhabi;$^{(b)}$United Arab Emirates University, Al Ain; United Arab Emirates.\\
% ...
$^{123}$Homer L. Dodge Department of Physics and Astronomy, University of Oklahoma, Norman OK; United States of America.\\
% ...
$^{155}$Raymond and Beverly Sackler School of Physics and Astronomy, Tel Aviv University, Tel Aviv; Israel.\\
% ...
$^{a}$ Also Affiliated with an institute covered by a cooperation agreement with CERN.\\
$^{b}$ Also at An-Najah National University, Nablus; Palestine.\\
$^{c}$ Also at Borough of Manhattan Community College, City University of New York, New York NY; United States of America.\\
$^{d}$ Also at Center for Interdisciplinary Research and Innovation (CIRI-AUTH), Thessaloniki; Greece.\\
$^{e}$ Also at Centro Studi e Ricerche Enrico Fermi; Italy.\\
$^{f}$ Also at CERN, Geneva; Switzerland.\\
% ...

\end{flushleft}
